\documentstyle [12pt,epsfig,rotating]{article}
\advance\topmargin by -2.8cm

\def\q2{$Q^2$}

\begin{document}
\hspace{10cm} CERN--PPE/95--138.

\hspace{10cm} hep-ph/9509406

\hspace{10cm} September, 4, 1995.
{}~
\vskip 1.3 cm
\begin{center}
{\LARGE\bf Measurement of the proton \\ and the
deuteron  structure functions, \\ $F_{2}^p$ and $F_{2}^d$} \\
\end{center}
\vskip 1 cm

\begin{center}
THE NEW MUON COLLABORATION (NMC) \\
\vspace{0.8cm}
{\footnotesize{\sl{
Bielefeld~University$^{1+}$,
Freiburg~University$^{2+}$,
Max-Planck~Institut f\"{u}r Kernphysik, Heidelberg$^{3+}$,
Heidelberg~University$^{4+}$,
Mainz~University$^{5+}$, Mons~University$^6$,
Neuch\^{a}tel~University$^7$,
NIKHEF$^{8++}$,
Saclay DAPNIA/SPP$^{9**}$,
University~of~California, Los Angeles$^{10}$,
University~of~California, Santa~Cruz$^{11}$,
Paul~Scherrer~Institut$^{12}$,
Torino~University and INFN~Torino$^{13}$,
Uppsala~University$^{14}$,
Soltan~Institute~for~Nuclear~Studies, Warsaw$^{15*}$,
Warsaw~University$^{16*}$}}}\\
\vspace{0.8cm}
{\small{M.~Arneodo$^{13a)}$,
A.~Arvidson$^{14}$,
B.~Bade{\l }ek$^{14,16}$,
M.~Ballintijn$^{8c)}$,
G.~Baum$^1$,
J.~Beaufays$^{8b)}$,
I.G.~Bird$^{3,8c)}$,
P.~Bj\"{o}rkholm$^{14}$,
M.~Botje$^{12d)}$,
C.~Broggini$^{7e)}$,
W.~Br\"{u}ckner$^3$,
A.~Br\"{u}ll$^{2f)}$,
W.J.~Burger$^{12g)}$,
J.~Ciborowski$^{8,16}$,
R.~van~Dantzig$^8$,
A.~Dyring$^{14}$,
H.~Engelien$^{2h)}$,
M.I.~Ferrero$^{13}$,
L.~Fluri$^7$,
U.~Gaul$^3$,
T.~Granier$^9$,
M.~Grosse-Perdekamp$^{2i)}$,
D.~von~Harrach$^{3j)}$,
M.~van~der~Heijden$^{8d)}$,
C.~Heusch$^{11}$,
G.~Igo$^{10}$,
Q.~Ingram$^{12}$,
M.~de~Jong$^8$,
E.M.~Kabu\ss$^{3j)}$,
R.~Kaiser$^2$,
H.J.Kessler$^2$,
T.J.~Ketel$^8$,
F.~Klein$^5$,
S.~Kullander$^{14}$,
U.~Landgraf$^2$,
T.~Lindqvist$^{14}$,
G.K.~Mallot$^{5}$,
C.~Mariotti$^{13l)}$,
G.~van~Middelkoop$^{8}$,
A.~Milsztajn$^9$,
Y.~Mizuno$^{3m)}$,
A.~Most$^{3w)}$,
A.~M\"{u}cklich$^3$,
J.~Nassalski$^{15}$,
D.~Nowotny$^{3n)}$,
J.~Oberski$^8$,
A.~Pai\'{c}$^7$,
C.~Peroni$^{13}$,
B.~Povh$^{3,4}$,
K.~Prytz$^{14v)}$,
R.~Rieger$^{5o)}$,
K.~Rith$^{3p)}$,
K.~R\"{o}hrich$^{5q)}$,
E.~Rondio$^{15}$,
L.~Ropelewski$^{16}$,
A.~Sandacz$^{15}$,
D.~Sanders$^{r)}$,
C.~Scholz$^{3n)}$,
R.~Seitz$^{5u)}$,
F.~Sever$^{1,8s)}$,
T.-A.~Shibata$^{4y)}$,
M.~Siebler$^1$,
A.~Simon$^{3t)}$,
A.~Staiano$^{13}$,
M.~Szleper$^{15}$,
Y.~Tzamouranis$^{3r)}$,
M.~Virchaux$^9$,
J.L.~Vuilleumier$^7$,
T.~Walcher$^5$,
C.~Whitten$^{10}$,
R.~Windmolders$^6$,
A.~Witzmann$^2$,
F.~Zetsche$^{3k)}$}} \\
\vspace{0.8cm}
\end{center}

\vspace{1cm}
\begin{center}
Submitted to Physics Letters
\end{center}

{\footnotesize {-----------------------------------\\

For footnotes see next page.}}
\newpage

\begin{tabbing}
{}~~~~\=+~~~\=Supported by Bundesministerium f\"{u}r Forschung und
Technologie.\\
\>	++\>	Supported in part by FOM, Vrije Universiteit Amsterdam and NWO.\\
\>	 *\>	Supported by KBN SPUB Nr 621/E - 78/SPUB/P3/209/94. \\
\>	**\>	Supported by CEA, Direction des Sciences de la Mati\a`ere.\\
\> \> \\
\>      a)\>    Now at Dipartimento di Fisica, Universit\a`a della Calabria, \\
\>        \>    87036 Arcavacata di Rende (Cosenza), Italy. \\
\>	b)\>	Now at Trasys, Brussels, Belgium.\\
\>	c)\>	Now at CERN, 1211 Gen\a`eve 23, Switzerland. \\
\>      d)\>    Now at NIKHEF, 1009 DB Amsterdam, The Netherlands. \\
\>      e)\>    Now at University of Padova, 35131 Padova, Italy.\\
\>      f)\>    Now at MPI f\"{u}r Kernphysik, 69029 Heidelberg, Germany. \\
\>	g)\>	Now at Universit\a'e de Gen\a`eve, 1211 Gen\a`eve 4, Switzerland.\\
\>	h)\>	Now at LHS GmbH, 63303 Dreieich, Germany.\\
\>      i)\>    Now at University of California, Los Angeles, 90024 Ca,
U.S.A.\\
\>      j)\>    Now at University of Mainz, 55099 Mainz, Germany. \\
\>      k)\>    Now at DESY, 22603 Hamburg, Germany.\\
\>	l)\>	Now at INFN-Istituto Superiore di Sanit\a`a, 00161 Roma, Italy.\\
\>	m)\>	Now at Osaka University, 567 Osaka, Japan.\\
\>	n)\>	Now at SAP AG, 69190 Walldorf, Germany.\\
\>      o)\>    Now at Comparex GmbH, 68165 Mannheim, Germany.\\
\>      p)\>    Now at University of Erlangen-N\"{u}rnburg, 91058 Erlangen,
Germany.\\
\>	q)\>	Now at IKP2-KFA, 52428 J\"{u}lich, Germany.\\
\>	r)\>	Now at University of Houston, 77204 TX, U.S.A.\\
\>      s)\>    Now at ESRF, 38043 Grenoble, France.\\
\>      t)\>    Now at New Mexico State University, Las Cruces NM, U.S.A.\\
\>      u)\>    Now at Dresden University, 01062 Dresden, Germany.\\
\>      v)\>    Now at Stockholm University, 113 85 Stockholm, Sweden.\\
\>      w)\>    Now University of Illinois, Illinois, USA. \\
\>      y)\>    Now University of Tokyo, Tokyo, Japan. \\

\end{tabbing}

\vskip 0.5 cm

{\large\bf	 Abstract}\\
\\
The proton and deuteron structure functions $F_2^p$ and $F_2^d$
were measured
in the kinematic range $0.006 < x < 0.6$ and
$0.5 < Q^2 < 75$ GeV$^2$,
by inclusive deep inelastic muon scattering
at 90, 120, 200 and 280~GeV.
The measurements are in good agreement with earlier  high
precision results.
The present and earlier results together have been parametrised
to give descriptions of the proton and deuteron structure functions
$F_2$ and their uncertainties over the range $0.006 < x < 0.9$.

\vskip 0.5 cm
\section{Introduction}
In this paper we present the
structure functions of the proton and the deuteron, $F_2^p$ and $F_2^d$,
obtained from deep
inelastic muon scattering at incident energies of 90, 120, 200 and 280 GeV.
The data at 90 GeV and half of the data at 280 GeV have already been published
\cite{f2nmc}. The present analysis supersedes that
of ref.\cite{f2nmc}.
The data cover the kinematic range $0.006 < x < 0.6$ and
$0.5 < Q^2 < 75$ GeV$^2$.

The structure function $F_2(x,Q^2)$ reflects the momentum
distribution of the quarks in the nucleon, an important aspect of
its internal structure.
The $Q^2$ dependence of $F_2$ can be used to determine the strong
coupling constant, $\alpha_s$,
and the momentum distribution of the gluons, whereas the energy dependence
of the cross section may
be used to extract $R(x,Q^2)$, the ratio of the longitudinal and transverse
virtual photon absorption cross sections.

The differential cross section for one-photon exchange can be written in
terms of the nucleon structure function $F_2(x,Q^2)$ and the ratio $R(x,Q^2)$,
as:
{\it
\begin{eqnarray}
\frac{{\rm d}^2\sigma(x,Q^2,E)}{{\rm d}x {\rm d} Q^2} &=&
\frac{4 \pi \alpha^2 }{Q^4} \cdot
\frac{F_2(x,Q^2)}{x} \cdot \nonumber \\
 && \cdot \left\{ 1 -y -\frac{Q^2}{4E^2}
+ (1 - \frac{2m^2}{Q^2}) \cdot \frac{y^2 +Q^2/E^2}
{2\left(1+R(x,Q^2)\right)} \right\},
\label{eq:sigma}
\end{eqnarray} }
where $\alpha$ is the fine structure constant, $-Q^2$ the four-momentum
transfer
squared, $E$ the energy of the incident muon
and $m$ the muon mass.
The quantities $x$, the Bjorken scaling variable, and $y$ are
defined as $x = Q^2/2M\nu$ and $y = \nu/E$,
where $\nu$
is the virtual photon energy in the laboratory frame and
$M$ is the proton mass. This one-photon exchange cross
section is obtained from the measured differential cross section by
correcting for higher order electroweak effects.

\section{The experiment}
The experiment (NMC--NA37) was performed at the M2 muon beam line
of the CERN SPS.
The data presented were taken in 1986 and 1987 at nominal incident energies of
90 and 280 GeV \cite{f2nmc}, and in 1989 at 120, 200 and 280 GeV.
The spectrometer is
described in refs. \cite{longratio,apparatus}.
The proton and deuteron differential
cross sections were measured simultaneously with two pairs of 3 m long
targets placed alternately in the muon beam. In one pair the upstream
target was liquid hydrogen and the downstream one
liquid deuterium, while in the other
pair the order was reversed. The spectrometer acceptance was substantially
different for the upstream and downstream targets, thereby giving two separate
measurements of the structure function for each material.

The integrated incident muon flux was measured both by sampling the beam
with a random trigger \cite{t10} and by recording the numbers of counts in two
scintillator hodoscope planes used to determine incident beam tracks
\cite{f2nmc}. In
both cases the beam tracks present in the triggers were reconstructed off-line,
in the same way as for scattered muon triggers, to determine the integrated
useable flux.

Uncertainties on the incident and scattered muon momenta
are important sources of systematic error in $F_2$. The
spectrometer which measured the scattered muon momentum (FSM)
was calibrated to an accuracy of 0.2\% using the reconstructed masses of
J/$\psi$ and K$^0$ mesons.
The beam momentum measurement system (BMS) was calibrated in dedicated
runs by measuring the average incident muon momentum in a purpose
built spectrometer \cite{bcs}.
The BMS was also calibrated relative to the FSM in a
series of runs using precision silicon microstrip detectors.
The two BMS calibrations were
averaged, leading to an accuracy of 0.2\%.

The following selections were applied to the data (see table \ref{tab:cutpar}).
The scattered muon momentum, $p'$, was  required to be above a
certain value to suppress muons from pion and kaon decays. Events with
small $\nu$, where the spectrometer resolution is poor, were rejected.
Regions
with rapidly varying acceptance were excluded by requiring minimum scattering
angles, $\theta_{min}$.
The maximum values of $y$ and the mass squared of the
hadronic final state, $W^2$, excluded the kinematic domain where higher order
electroweak processes dominate. In addition,
the position of the reconstructed vertex was constrained to be within one of
the targets.
At  each value of $x$, data in regions of $Q^2$ where the
acceptance was less than 30\% of the maximum at that $x$ were removed.
Inefficient regions in the spectrometer were excluded.
As in the previous analysis \cite{f2nmc}, the tracks reconstructed in
the large drift chambers (W45) in front of the absorber
were not used in the analysis.

\section{Data analysis and results}
The extraction of the structure functions
from the measured cross sections
requires the evaluation of the
higher order electroweak contributions and thus
the a priori knowledge of the structure functions themselves.
Also, the correction for the effects of kinematic smearing
requires the knowledge of the event distributions
in the $(x,Q^2)$ plane
and thus the knowledge of the cross sections.
Therefore, the structure functions were determined iteratively.

A Monte Carlo simulation of the experiment was performed.
At each iteration, the Monte Carlo events were weighted with the best
knowledge of the total cross section, i.e. the one-photon
exchange cross section
computed from parametrisations of $F_2$ and $R$, corrected for
higher order electroweak processes.
By comparing the Monte Carlo simulation with the
data, a new parametrisation of $F_2$ was determined for each target material.
Convergence was considered to be attained when the values of $F_2$
changed by less than 0.2~\%, in practice after two iterations.
A phenomenological 15-parameter function, given in eq.
(\ref{eq:parF2}) below, was used to parametrise $F_2$.
The parametrisation of $R$ was
taken from ref. \cite{rslac} and kept fixed in the iteration.

Separate Monte Carlo simulations were performed for each period of
data taking. This enabled changes in the beam
and the detector to be taken into account.
The simulation of the experiment
was checked by comparing the distributions of data and Monte Carlo
events in variables not or only weakly related to $x$ and $Q^2$.

As in ref.\cite{f2nmc} the uncertainty
in the determination of the spectrometer acceptance
was studied by comparing
the structure functions determined separately from the upstream and
the downstream targets, for which the spectrometer has largely
different acceptances.
In addition, this uncertainty was studied
by comparing the structure function results  obtained using two
different methods.
In the first one a common parametrisation was used to
determine the structure function for all energies
simultaneously. In the second method the cross sections were
initially extracted for each data period
and target position
separately, and afterwards combined to determine the
structure function.
These studies led to an estimated contribution to the
systematic error on $F_2$
of the order of 2\%, reaching 4\% at the
edge of the kinematic domain.
A further systematic error, of up to 2\% in the
small $x$ and large $y$ region, was attributed to $F_2$ to
account for uncertainties in the effect of hadronic
and electromagnetic showers on the reconstruction efficiency.

The higher order electromagnetic contributions to the cross section
and their uncertainties
were calculated as in refs. \cite{f2nmc,longratio}
using the method of Akhundov, Bardin and Shumeiko \cite{radcor}.
These contributions to the cross sections were
at most one third.
The consequent systematic errors on $F_2$ arise
predominantly from the uncertainties
in $R$, in the
proton form factor and in the suppression
of the quasi-elastic scattering on the deuteron. The uncertainty
in $R$ contributes both here and
through the calculation
of the one-photon cross section (eq.~(\ref{eq:sigma})).

The normalisation uncertainty of the data at each incident
energy, relative to the fitted function describing
$F_2$ used in the iteration, is estimated to be 2\%.
This is included in the 2.5\% total normalisation uncertainty of
the combined data.

\begin{figure}[p]
\begin{sideways}
 \begin{minipage}[b]{\textheight}
 \begin{center}
 \begin{tabular}{c c}
 \epsfig{figure=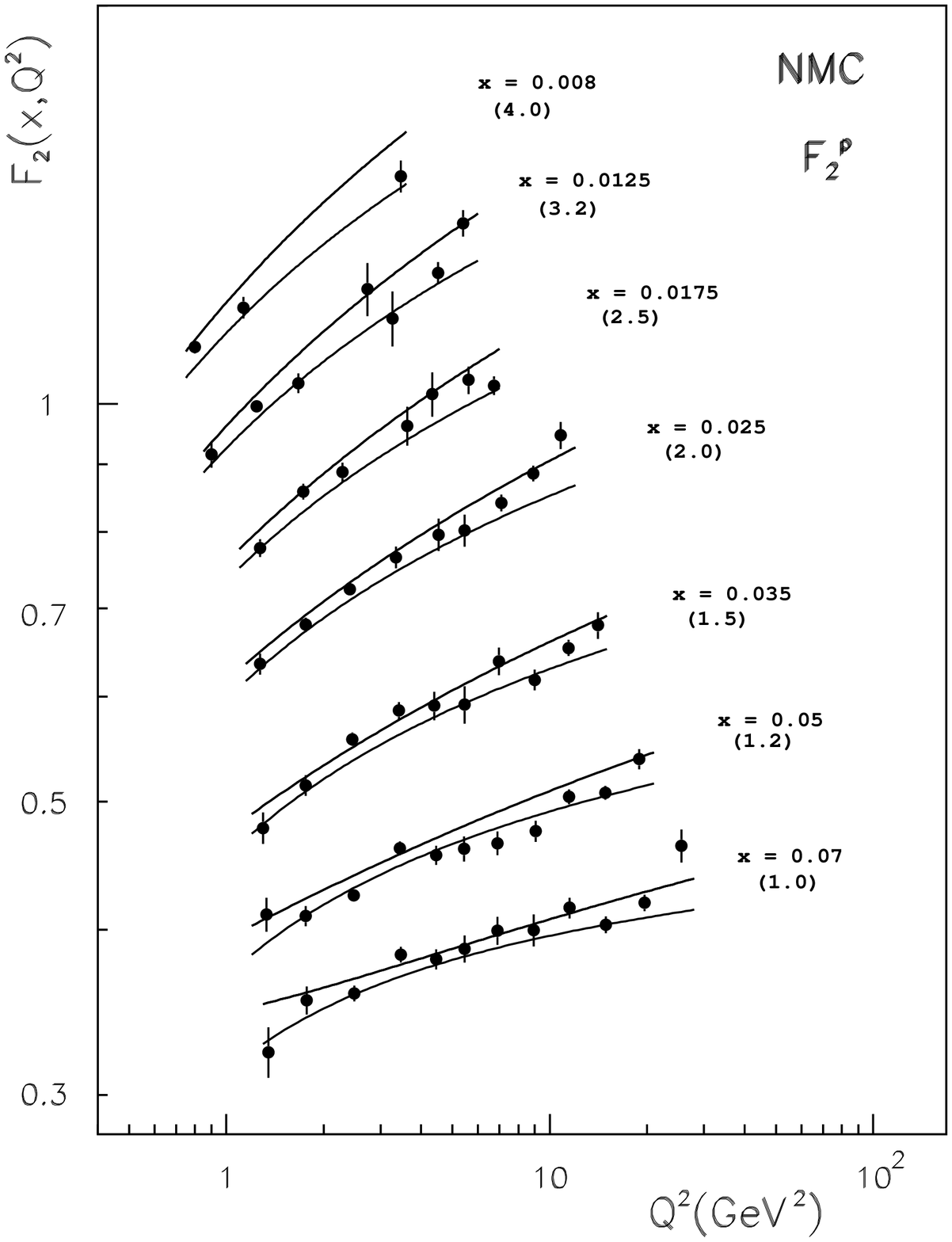,height=12cm,width=11cm} &
 \epsfig{figure=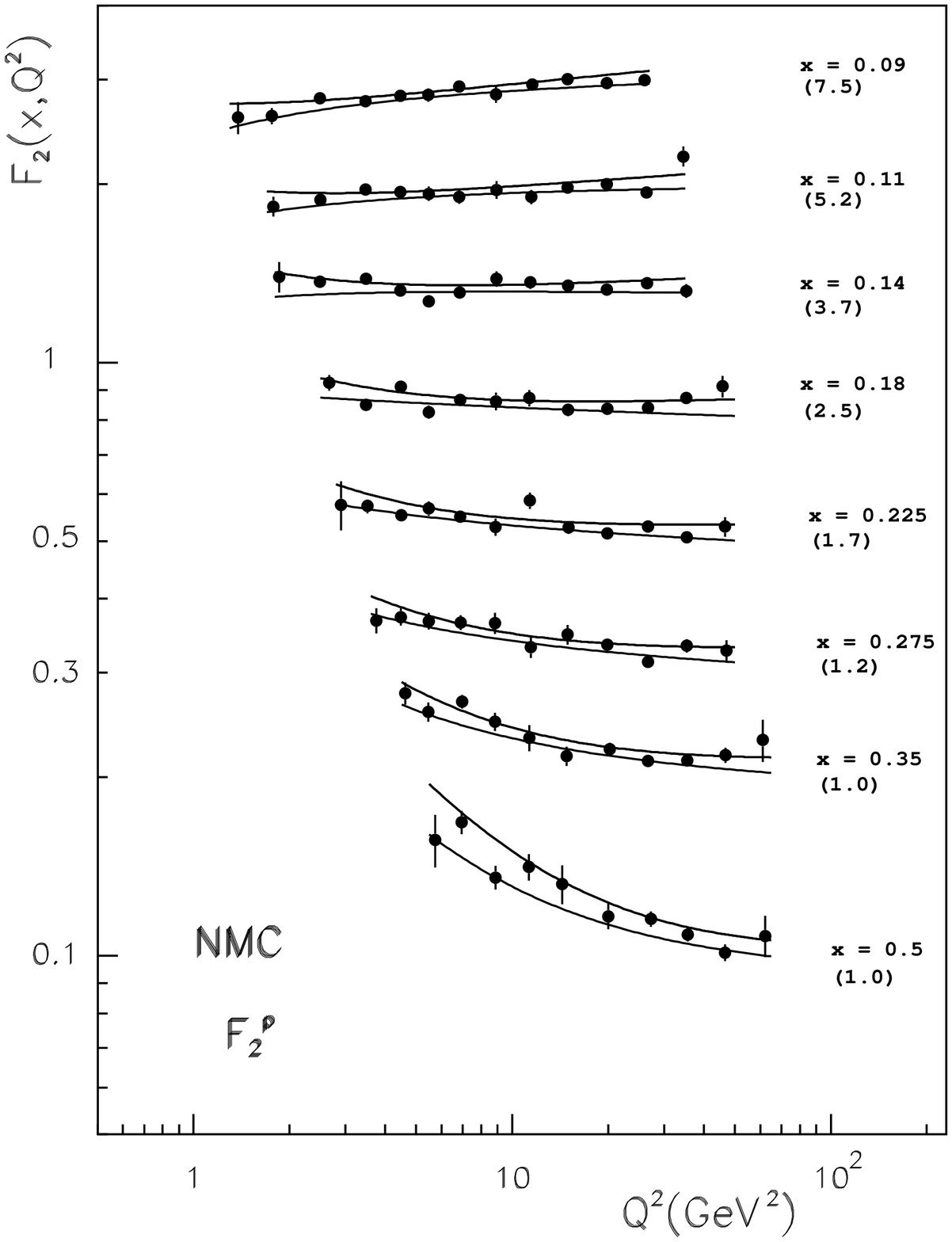,height=12cm,width=11cm}
 \end{tabular}
 \end{center}
\caption{
The proton structure function $F_2^p$.
In the figure the data in each $x$ bin have been scaled by the
factors indicated in brackets for clarity.
The error bars
represent the statistical uncertainties, the solid lines the
systematic uncertainties.
}
\label{fig:NMCprot}
\end{minipage}
\end{sideways}
\end{figure}

\begin{figure}[p]
\begin{sideways}
 \begin{minipage}[b]{\textheight}
 \begin{center}
 \begin{tabular}{c c}
 \epsfig{figure=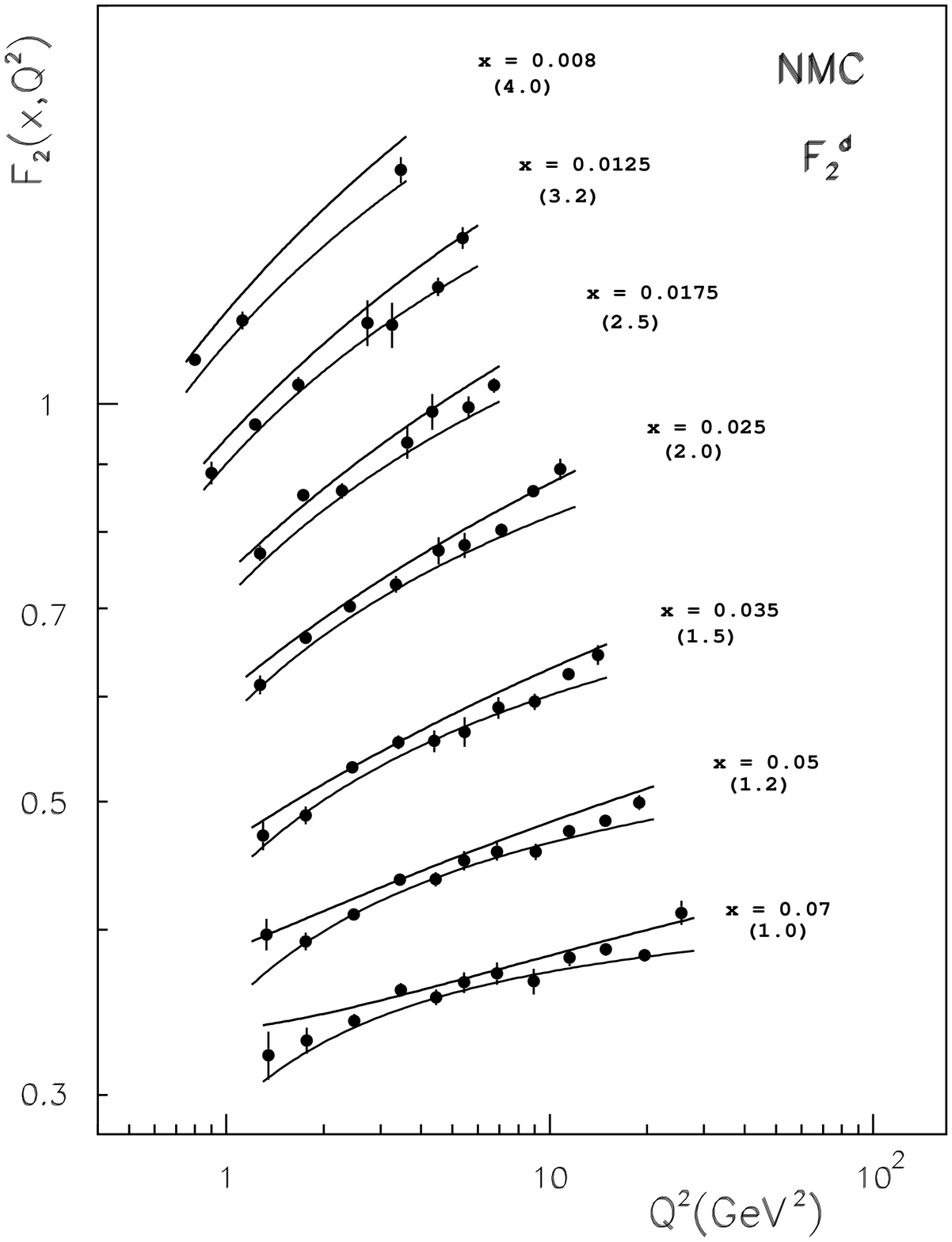,height=12cm,width=11cm} &
 \epsfig{figure=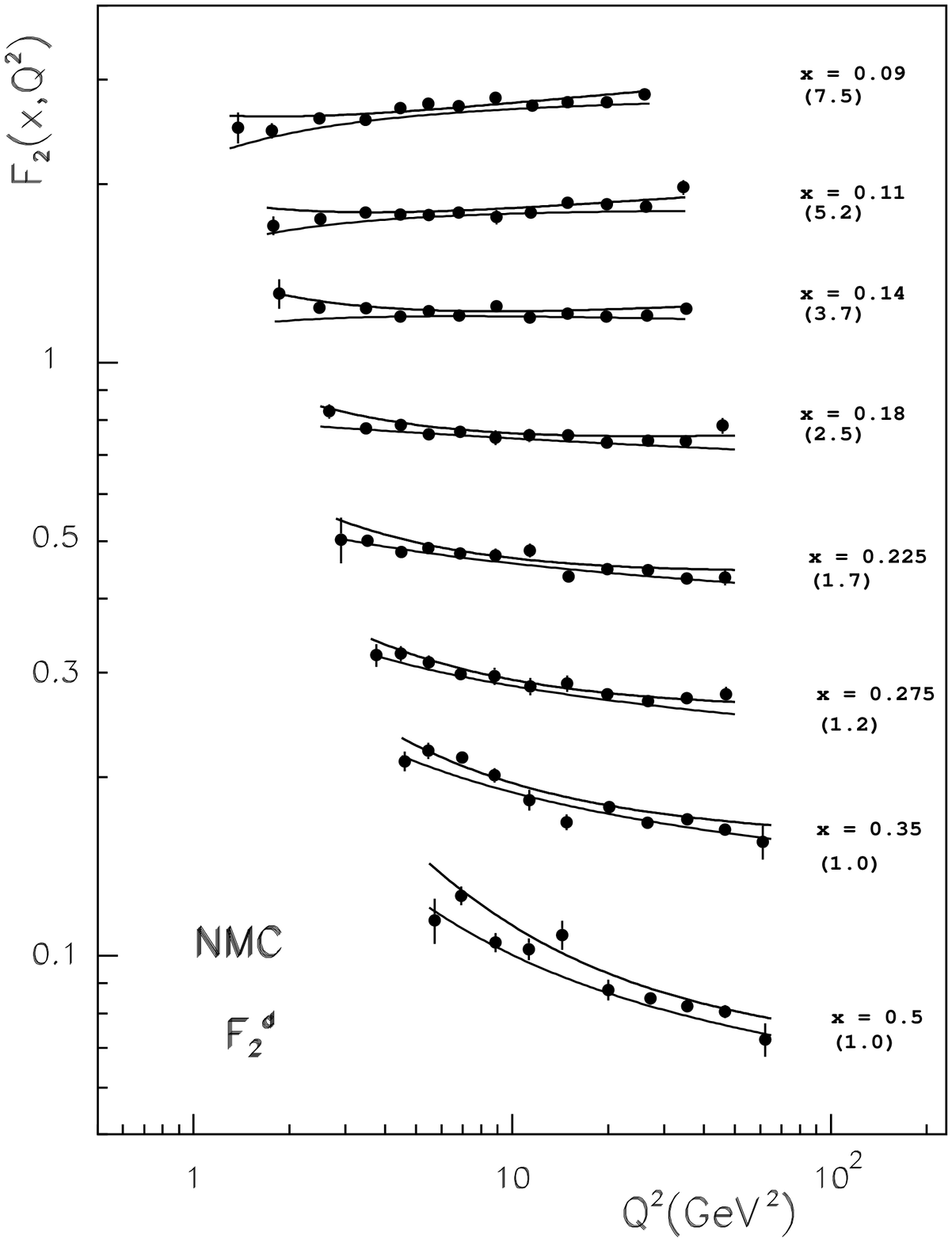,height=12cm,width=11cm}
 \end{tabular}
 \end{center}
\caption{
The deuteron structure function $F_2^d$.
In the figure the data in each $x$ bin have been scaled by the
factors indicated in brackets for clarity.
The error bars
represent the statistical uncertainties, the solid lines the
systematic uncertainties.
}
\label{fig:NMCdeut}
\end{minipage}
\end{sideways}
\end{figure}

The results obtained for the structure functions $F_2^p$ and $F_2^d$ averaged
over all energies are presented as a function of \q2 for fixed values of $x$
in figs.~\ref{fig:NMCprot} and \ref{fig:NMCdeut}.
The values of $F_2^p(x,Q^2)$ and $F_2^d(x,Q^2)$  and the various contributions
to the systematic errors are given in tables \ref{tab:h2} and \ref{tab:d2}.
In figs.~\ref{fig:NMCprot} and \ref{fig:NMCdeut}
the error bars represent the statistical errors, while the quadratic sum of the
various systematic uncertainties is given by the lines,
which
are drawn relative to the function fitted to the data and
do not include any normalisation uncertainty.

\begin{figure}[p]
 \epsfig{figure=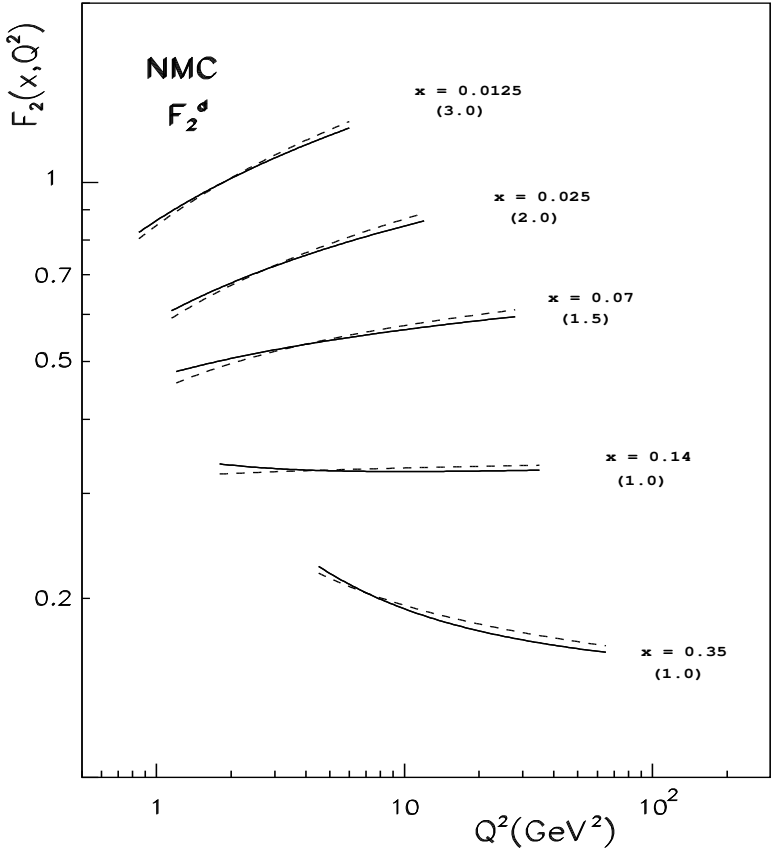,height=12cm,width=11cm}
\caption{
The effect of changes in the relative normalisations shown for five
$x$ bins.
The solid lines represent the function
fitted to the $F_2$ results, while
the dashed lines show a similar fit with
the 90~GeV data lowered by 2\% and the other three data sets
raised by 2\%.
 }
\label{fig:relnorm}
\end{figure}

An indication of the uncertainty in the \q2 dependence of
$F_2$ due to the relative normalisation uncertainties is
illustrated in fig.~\ref{fig:relnorm}. Here the function
fitted to the $F_2$ results is compared to a similar fit with
the 90~GeV data lowered by 2\% and the other three data sets
raised by 2\%.

\section{Discussion}
Our results are in good agreement with those of SLAC
\cite{slac92} and BCDMS \cite{BCDMS}.
We have used the results of these experiments and the present one to
obtain  parametrisations of the structure functions $F_2^p$
and $F_2^d$ and
their uncertainties,
using the 15-parameter function \cite{15param}:

\begin{equation}
{\it
F_2(x,Q^2)  =  A(x)\cdot(\frac{\ln(Q^2/\Lambda^2)}
   {\ln({Q_0}^2/\Lambda^2)})^{B(x)}
   \cdot (1+\frac{C(x)}{Q^2}),}
\label{eq:parF2}
\end{equation}
with
$Q_0^2  =  20$ GeV$^2$, $\Lambda=0.250\: $GeV  and\\
{\it
$A(x)  =
x^{a_1}\cdot(1-x)^{a_2}\cdot\{a_3+a_4(1-x)+a_5(1-x)^2+a_6(1-x)^3+
   a_7(1-x)^4\},$\\
$B(x)  =  b_1+b_2x+b_3/(x+b_4),$\\
$C(x)  =  c_1x+c_2x^2+c_3x^3+c_4x^4.$\\}

In the fits, the individual data points were weighted using their
statistical errors only.
Five additional normalisation parameters were fitted to the
data to describe
relative normalisation shifts between the four
NMC data sets taken at different energies
and the SLAC and the BCDMS data, weighted according to their
normalisation uncertainties
(2\% for the four NMC sets, 2\% for the SLAC data and 3\% for the BCDMS
results).
Optimal agreement between the experiments was found
with the changes in normalisation given in table \ref{tab:normpar}.
An additional free parameter was included to account for a
possible miscalibration of the scattered muon energy in the BCDMS data
and was determined from the fit
to be $+$0.2\% for both the proton and the deuteron data.
The parameters of eq.~(\ref{eq:parF2}) resulting from the fits
are given in table~\ref{tab:f2par}.

To determine the total uncertainties in the parametrised
structure functions $F_2^p$ and $F_2^d$, we also took the systematic errors and
their correlations into account.
The total errors on $F_2^p$ and $F_2^d$
are between 1.5\% ~and~ 5\% and include the normalisation uncertainty.
The upper and lower limits of the structure functions $F_2$ were fitted with
the function
of eq.~(\ref{eq:parF2}) and the resulting parameter values are listed  in
table~\ref{tab:f2par}.

\begin{figure}[p]
\begin{sideways}
 \begin{minipage}[b]{\textheight}
 \begin{center}
 \begin{tabular}{c c}
 \epsfig{figure=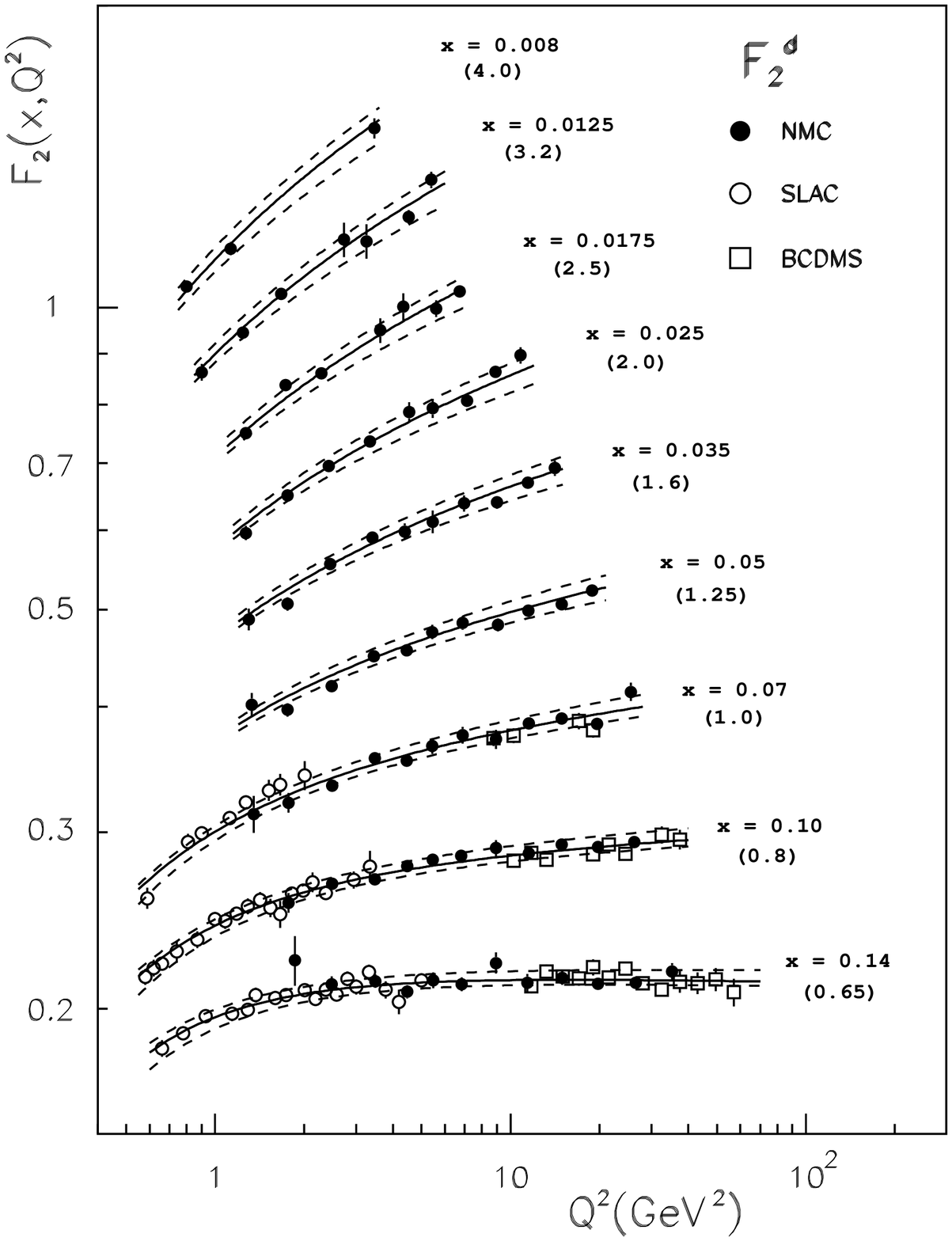,height=12cm,width=11cm} &
 \epsfig{figure=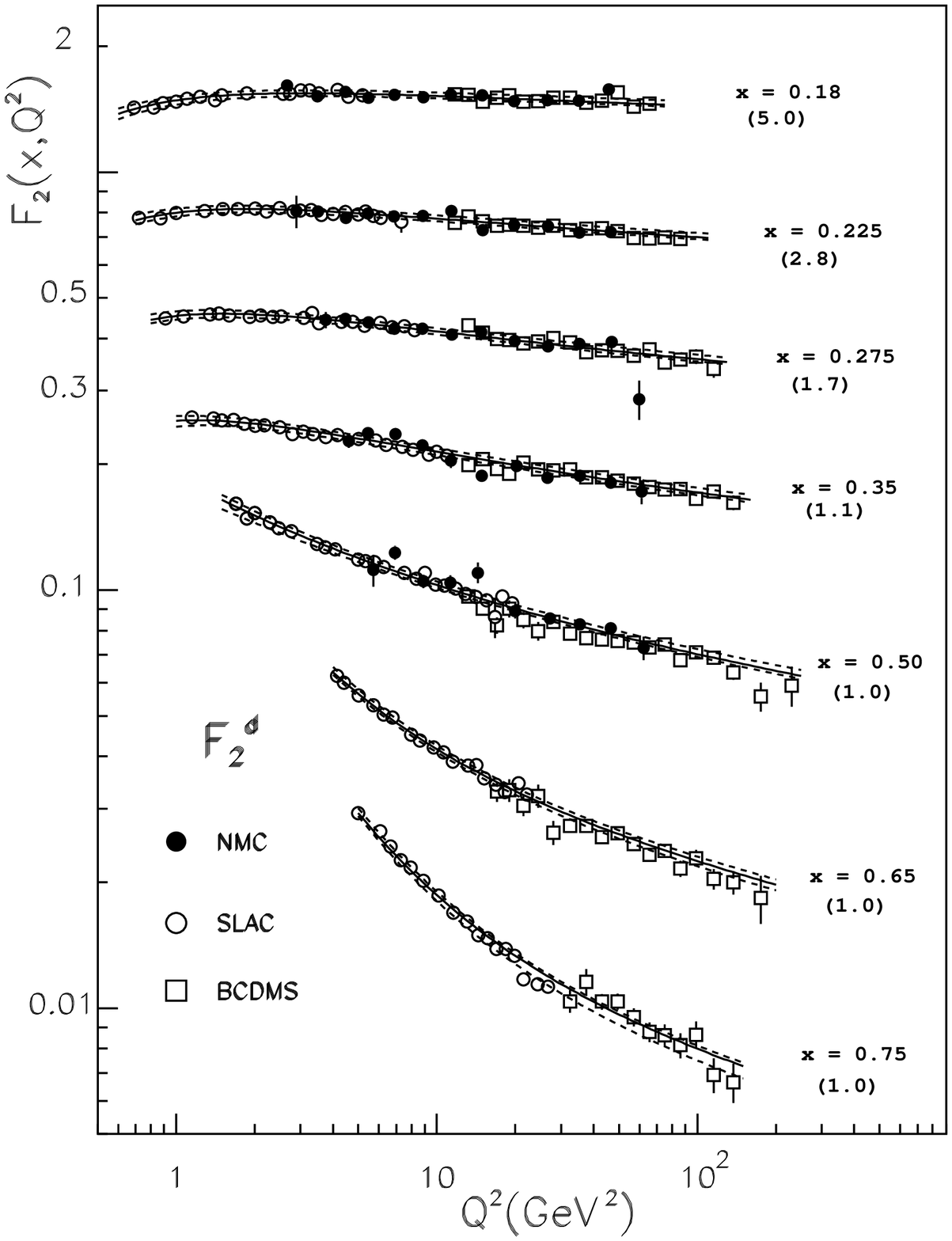,height=12cm,width=11cm}
 \end{tabular}
 \end{center}
\caption{
The deuteron structure function $F_2^d$.
The NMC results are
compared with those of SLAC \protect{\cite{slac92}} and BCDMS
\protect{\cite{BCDMS}}.
The points have been renormalised
according to the values resulting from the fit,
and the BCDMS data have also been adjusted for the energy recalibration
obtained from the fits.
The SLAC and BCDMS values were rebinned to the NMC $x$ bins.
The error bars
represent the statistical errors.
The solid curves are the result of the fit of the 15-parameter function
(eq.~(\protect{\ref{eq:parF2}})) to the three data sets.
The dashed curves indicate the total uncertainty.
The data in each $x$ bin are scaled by the
factors indicated in brackets for clarity.
}
\label{fig:fitdeut}
\end{minipage}
\end{sideways}
\end{figure}

The result of the fit for the deuteron is shown
in fig.~\ref{fig:fitdeut} together
with the merged NMC data and the SLAC and BCDMS results.
In the figure, the points have been renormalised
according to the values resulting from the fit,
and the BCDMS data have also been adjusted for the 0.2\%
energy recalibration mentioned above.

\begin{figure}[p]
\begin{sideways}
 \begin{minipage}[b]{\textheight}
 \epsfig{figure=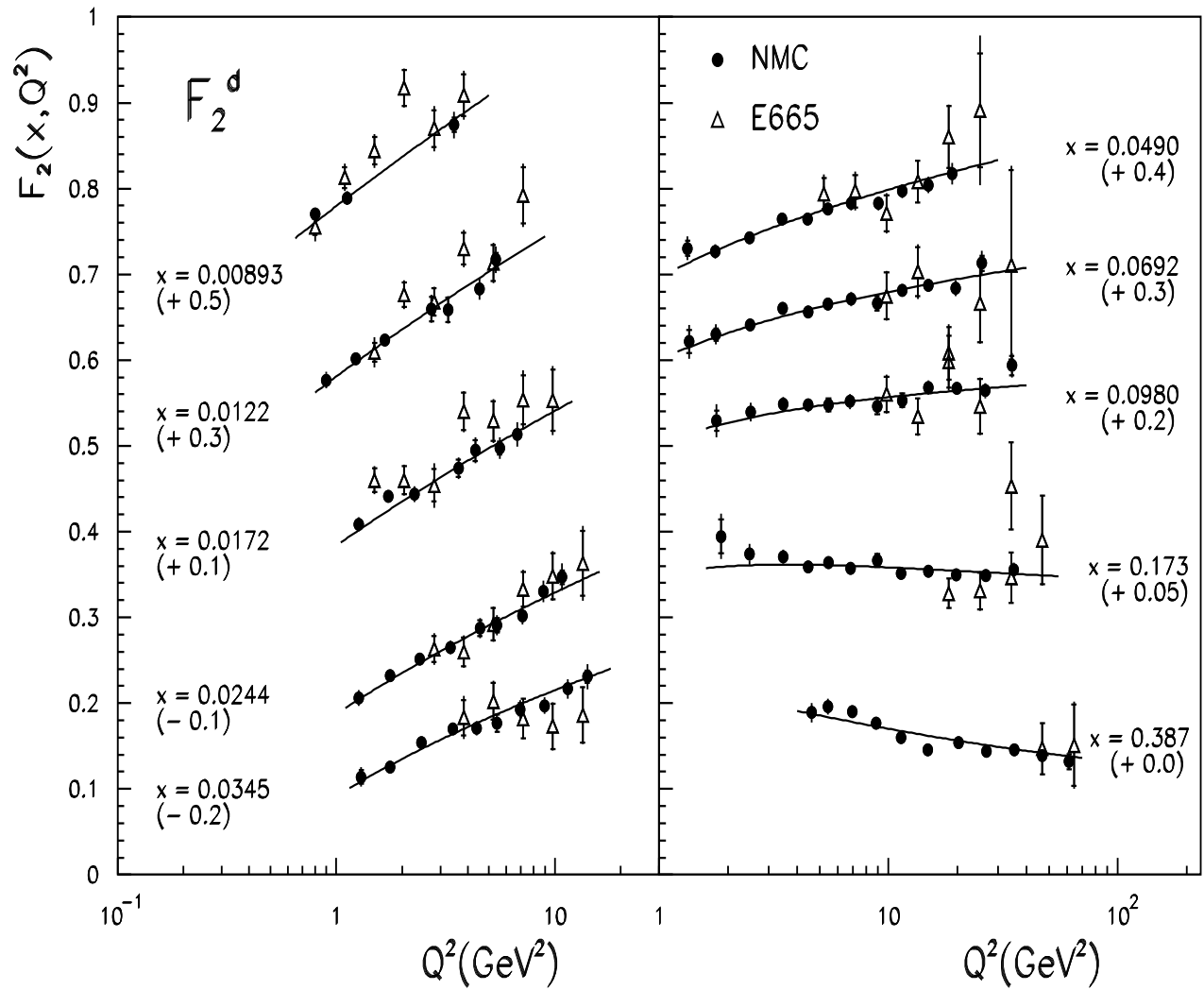,height=15cm,width=18cm}
\caption{
Comparison of the NMC and E665 \protect{\cite{ref:E665}} results for $F_2^d$.
The data of NMC were
interpolated to the $x$ bins of the E665 data using the
current parametrisation of $F_2^d$. The inner
error bars indicate the statistical errors, the outer bars the
quadratic sum of the statistical and systematic errors.
The lines correspond to the $F_2$ parametrisation, as
shown in fig. \protect{\ref{fig:fitdeut}}.
The data in each $x$ bin are offset by the amounts indicated
in brackets.
}
\label{fig:E665}
\end{minipage}
\end{sideways}
\end{figure}

Recently, preliminary structure function results became available from
the fixed target experiment E665
at Fermilab \cite{ref:E665}. These data extend to very small $x$ (0.0004).
In fig.~\ref{fig:E665} our data are compared
with the E665 results in the overlapping
$x$ region. The agreement is generally good, although for
small $x$ there is a tendency for these data to be above
those presented here.

\begin{figure}[p]
\begin{sideways}
 \begin{minipage}[b]{\textheight}
 \epsfig{figure=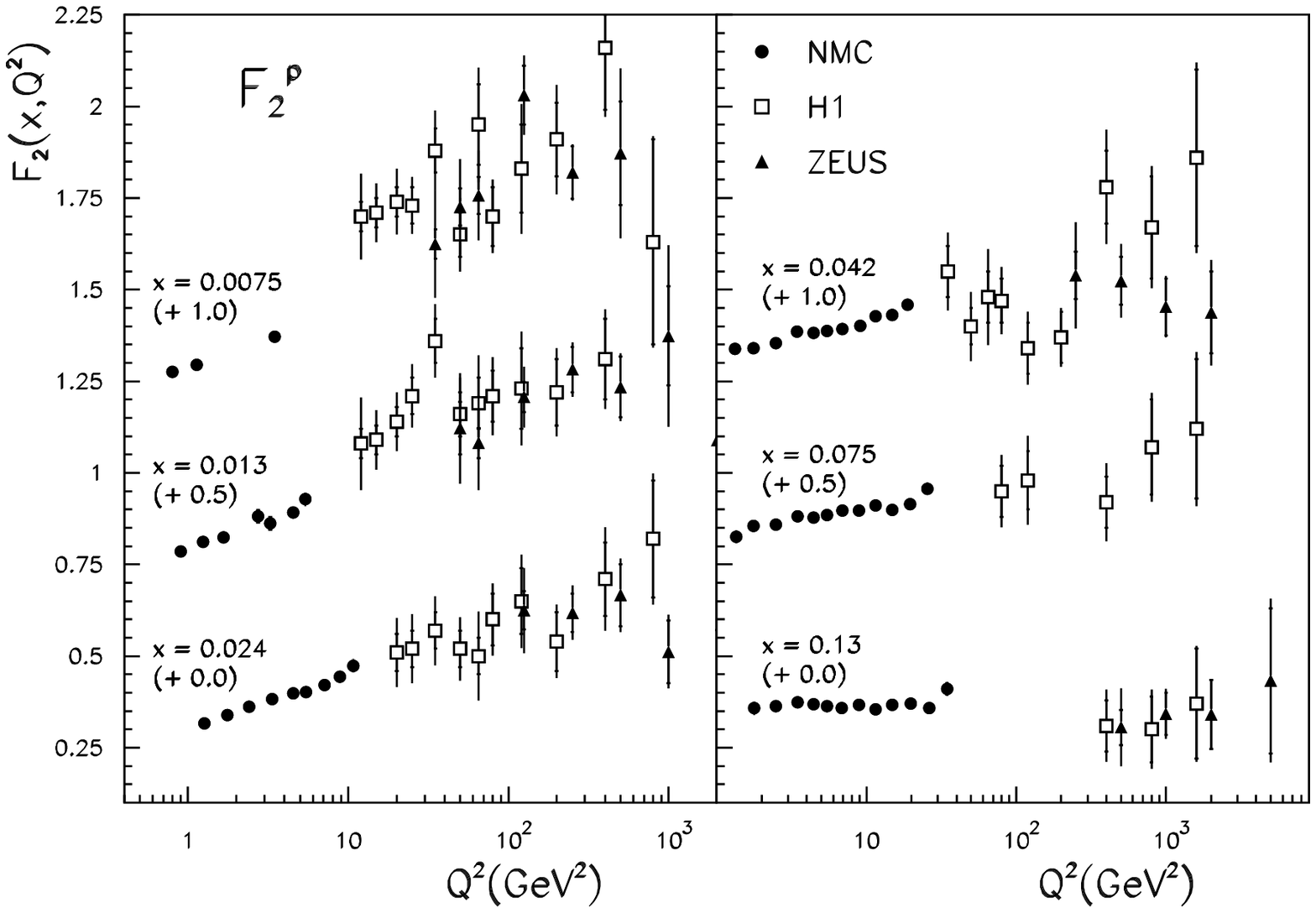,height=15cm,width=20cm}
\caption{
Comparison of the NMC and HERA results for $F_2^p$.
The data of NMC and ZEUS data were interpolated
to the $x$ bins of H1 \protect{\cite{ref:h1}},
using the current parametrisation for the NMC data and that of ref.
\protect{\cite{ref:zeus}} for the ZEUS data. The inner
error bars indicate the statistical errors, the outer bars the
quadratic sum of the statistical and systematic errors.
The data in each $x$ bin are offset by the amounts indicated
in brackets.
}
\label{fig:H1}
\end{minipage}
\end{sideways}
\end{figure}

In fig.~\ref{fig:H1} we present a comparison
of the NMC data with the results published by
the H1 \cite{ref:h1}
and ZEUS \cite{ref:zeus} collaborations at HERA.
Again, the agreement is good, except for the lowest $x$ bin.

{\vskip 5mm}

\section{Summary}

We have presented the
proton and deuteron structure functions $F_2^p$ and $F_2^d$
in the kinematic range $0.006 < x < 0.6$ and
$0.5 < Q^2 < 75$ GeV$^2$,
obtained from inclusive deep inelastic muon scattering experiments
at 90, 120, 200 and 280~GeV.
The combined results have high statistical accuracy and the
systematic uncertainties are between 2\% and 5\%.
The data are in good agreement with the results from
SLAC and BCDMS.
Agreement is also observed with the Fermilab E665 data, while
published data on $F_2$ from the HERA experiments H1 and ZEUS
are consistent
with the present results for $x \geq 0.01$.

The present results together with the earlier high precision
data of SLAC and BCDMS
were parametrised to give descriptions of the
proton and deuteron structure functions $F_2$
and their uncertainties over the range $0.006 < x < 0.9$.

\clearpage
\begin{table}[p]
\centering
\begin{tabular}{ccccccccc} \hline
 Incident energy & $p'_{min}$ & $\nu_{min}$ & $\theta^{up}_{min}$ &
 $\theta^{down}_{min}$ & $y_{max}$ & $W^2_{max}$ & $N_p$ & $N_d$ \\
 ~[GeV]   & [GeV]      & [GeV] & [mrad]              &
[mrad]                &     & [GeV$^2$] & [$10^3$]  &  [$10^3$] \\
 \hline
 90 & 15 &  7 & 13 & 15 & 0.9 & 130 & 255 & 533 \\
120 & 20 & 10 & 13 & 15 & 0.9 & 150 & 103 & 215 \\
200 & 35 & 20 & 13 & 15 & 0.9 & 250 &  56 & 114 \\
280 & 40 & 30 & 13 & 15 & 0.9 & 400 & 141 & 406 \\
\hline
\end{tabular}
\caption{Cuts applied to the data, as explained in the text.
Different values of $\theta_{min}$ were used for the upstream and downstream
targets, as indicated. $N_p$ and $N_d$ are the total number of events for
protons and deuterons, respectively, after applying all cuts.
}
\label{tab:cutpar}
\end{table}

\begin{table}[p]
\centering
\begin{tabular}{lrr}
\hline
Data set & Proton   & Deuteron         \\
\hline
SLAC         &  +0.1\%    & +0.4\%      \\
NMC  90 GeV  &  --2.2\%   & --2.7\%   \\
NMC 120 GeV  &  +1.6\%    & +1.4\%   \\
NMC 200 GeV  &  +1.2\%    & +1.6\%   \\
NMC 280 GeV  &  +0.8\%    & +0.5\%   \\
BCDMS        &  --2.3\%    & --1.9\%   \\
\hline
\end{tabular}
\caption{Normalisation changes for the different data sets.
All numbers were obtained from the fits described in section~4.
}
\label{tab:normpar}
\end{table}

\begin{table}[p]
\footnotesize
\centering
\begin{tabular}{ccccccc}
\hline
             &             & \multicolumn{2}{c}{Limits} &
                           & \multicolumn{2}{c}{Limits}  \\
Parameter    &  $F_2^p$    & Upper  $F_2^p$ & Lower  $F_2^p$ &
                $F_2^d$    & Upper  $F_2^d$ & Lower  $F_2^d$ \\
\hline
$a_1$        &  -0.02778  &  -0.05711  &  -0.01705  & -0.04858 & -0.04715  &
-0.02732  \\
$a_2$        &  2.926  &  2.887  &  2.851  &  2.863  & 2.814  &  2.676  \\
$a_3$        &  1.0362  &  0.9980  &  0.8213  &  0.8367 & 0.7286  &  0.3966  \\
$a_4$        &  -1.840  &  -1.758  &  -1.156  &  -2.532 & -2.151  &  -0.608  \\
$a_5$        &  8.123  &  7.890  &  6.836  &  9.145  & 8.662  &  4.946  \\
$a_6$        &  -13.074  &  -12.696  &  -11.681  &  -12.504  & -12.258  &
-7.994  \\
$a_7$        &  6.215  &  5.992  &  5.645  &  5.473  & 5.452  &  3.686  \\
$b_1$        &  0.285  &  0.247  &  0.325  &  -.008 & -.048  &  0.141  \\
$b_2$        &  -2.694  &  -2.611  &  -2.767 &  -2.227  &  -2.114  &  -2.464
\\
$b_3$        &  0.0188  &  0.0243  &  0.0148 &  0.0551  &  0.0672  &  0.0299
\\
$b_4$        &  0.0274  &  0.0307  &  0.0226 &  0.0570  &  0.0677  &  0.0396
\\
$c_1$        &  -1.413  &  -1.348  &  -1.542 &  -1.509  &  -1.517  &  -2.128
\\
$c_2$        &   9.366  &   8.548  &  10.549 &  8.553   &   9.515  &  14.378
\\
$c_3$        &   -37.79  &  -35.01  &   -40.81  &  -31.20 &   -34.94  &
-47.76  \\
$c_4$        &    47.10  &    44.43  &    49.12  &  39.98  &    44.42  &
53.63  \\
\hline
\end{tabular}
\caption{The values of the parameters of  eq.~(\protect{\ref{eq:parF2}})
for $F_2^p$ and $F_2^d$.
The limits correspond to the total uncertainties in $F_2$
as explained in the text.
}
\label{tab:f2par}
\end{table}

\clearpage

\newpage
\textwidth=15cm
\footnotesize
\begin{table}[t]                                                                                                                   
~
\vspace{1cm}
\caption{
The proton structure function $F_2^p(x,Q^2)$ from the measurements at 90, 120,
200 and 280~GeV, averaged over all energies.
In this table the value of $F_2$ and its statistical and
total systematic errors are given for each $x$ and $Q^2$.
The mean $y$ of the events in this bin is also given.
Also listed is the measured cross section (${\rm d}^2\sigma^{meas}/
{\rm d}x{\rm d}Q^2$) at the same $x$ and $Q^2$.
The systematic error is the quadratic sum of the contributions given as
percentages of $F_2$ in the columns 4--8.  The contributions are:
E, E', errors due to calibrations of
incident and scattered muon energies;
AC, error in the determination of the spectrometer acceptance;
RC, error from the radiative corrections;
RE, error from the reconstruction inefficiency due to showers.
}
\label{tab:h2}
\vspace{1cm}
 
\begin{tabular}{|c|c|c||c|c|c|c|c||c|r|}\hline
$x$ & $Q^2$ & $y$ & $E$ & $E^{\prime}$ & $AC$ & $RC$ & $RE$& $d^2\sigma^{meas}/dxdQ^2$ &
$F^p_2\;\pm\Delta{F^{stat}_2}\;\pm\Delta{F^{syst}_2}$ \\
  & $[{\rm GeV^2}]$ & &  [\%]  &  [\%]  &  [\%]  &  [\%]  &  [\%]  & $[{\rm b\cdot GeV^{-2}}]$ & \\ 
 \hline
  0.0080 &   0.800 &   0.617 &  0.6 &  0.1 &  2.3 &  1.7 &  0.3 & 0.928E-05 &  0.2760 $\pm$  0.0031 $\pm$  0.0081 \\
  0.0080 &   1.130 &   0.744 &  0.5 &  0.0 &  2.1 &  1.8 &  0.5 & 0.495E-05 &  0.2955 $\pm$  0.0055 $\pm$  0.0085 \\
  0.0080 &   3.470 &   0.855 &  0.1 &  0.5 &  2.1 &  1.8 &  2.0 & 0.747E-06 &  0.3715 $\pm$  0.0103 $\pm$  0.0128 \\
 \hline
  0.0125 &   0.900 &   0.444 &  0.7 &  0.3 &  1.6 &  0.8 &  0.2 & 0.546E-05 &  0.2859 $\pm$  0.0064 $\pm$  0.0056 \\
  0.0125 &   1.240 &   0.582 &  0.6 &  0.1 &  1.6 &  1.3 &  0.3 & 0.280E-05 &  0.3112 $\pm$  0.0034 $\pm$  0.0068 \\
  0.0125 &   1.670 &   0.665 &  0.5 &  0.1 &  1.5 &  1.3 &  0.5 & 0.159E-05 &  0.3238 $\pm$  0.0055 $\pm$  0.0069 \\
  0.0125 &   2.730 &   0.604 &  0.6 &  0.1 &  1.5 &  1.1 &  1.2 & 0.675E-06 &  0.3816 $\pm$  0.0177 $\pm$  0.0087 \\
  0.0125 &   3.260 &   0.710 &  0.5 &  0.0 &  1.7 &  1.2 &  1.3 & 0.471E-06 &  0.3626 $\pm$  0.0172 $\pm$  0.0092 \\
  0.0125 &   4.520 &   0.714 &  0.1 &  0.5 &  2.0 &  2.3 &  1.7 & 0.260E-06 &  0.3924 $\pm$  0.0076 $\pm$  0.0137 \\
  0.0125 &   5.390 &   0.839 &  0.1 &  0.5 &  2.2 &  2.5 &  1.9 & 0.201E-06 &  0.4279 $\pm$  0.0099 $\pm$  0.0167 \\
 \hline
  0.0175 &   1.270 &   0.444 &  0.8 &  0.2 &  1.1 &  0.7 &  0.2 & 0.213E-05 &  0.3111 $\pm$  0.0049 $\pm$  0.0048 \\
  0.0175 &   1.730 &   0.557 &  0.6 &  0.1 &  1.2 &  0.9 &  0.3 & 0.112E-05 &  0.3432 $\pm$  0.0047 $\pm$  0.0057 \\
  0.0175 &   2.290 &   0.635 &  0.5 &  0.0 &  1.1 &  1.0 &  0.5 & 0.641E-06 &  0.3550 $\pm$  0.0060 $\pm$  0.0060 \\
  0.0175 &   3.620 &   0.567 &  0.6 &  0.1 &  1.5 &  1.2 &  1.1 & 0.290E-06 &  0.3848 $\pm$  0.0129 $\pm$  0.0087 \\
  0.0175 &   4.330 &   0.674 &  0.5 &  0.0 &  1.6 &  1.7 &  1.3 & 0.198E-06 &  0.4067 $\pm$  0.0157 $\pm$  0.0112 \\
  0.0175 &   5.600 &   0.632 &  0.1 &  0.5 &  1.9 &  2.2 &  1.5 & 0.126E-06 &  0.4168 $\pm$  0.0100 $\pm$  0.0137 \\
  0.0175 &   6.730 &   0.749 &  0.1 &  0.5 &  2.1 &  2.3 &  1.7 & 0.877E-07 &  0.4128 $\pm$  0.0066 $\pm$  0.0146 \\
 \hline
  0.0250 &   1.270 &   0.315 &  1.0 &  0.5 &  0.4 &  0.6 &  0.1 & 0.166E-05 &  0.3177 $\pm$  0.0058 $\pm$  0.0043 \\
  0.0250 &   1.760 &   0.415 &  0.8 &  0.3 &  0.7 &  0.5 &  0.2 & 0.843E-06 &  0.3401 $\pm$  0.0042 $\pm$  0.0041 \\
  0.0250 &   2.410 &   0.517 &  0.7 &  0.1 &  0.8 &  0.7 &  0.3 & 0.433E-06 &  0.3620 $\pm$  0.0035 $\pm$  0.0045 \\
  0.0250 &   3.340 &   0.628 &  0.6 &  0.0 &  1.0 &  1.2 &  0.5 & 0.223E-06 &  0.3825 $\pm$  0.0072 $\pm$  0.0066 \\
  0.0250 &   4.540 &   0.500 &  0.7 &  0.1 &  1.3 &  1.2 &  1.0 & 0.138E-06 &  0.3979 $\pm$  0.0112 $\pm$  0.0085 \\
  0.0250 &   5.450 &   0.595 &  0.6 &  0.0 &  1.5 &  1.6 &  1.1 & 0.919E-07 &  0.4009 $\pm$  0.0112 $\pm$  0.0101 \\
  0.0250 &   7.090 &   0.587 &  0.1 &  0.0 &  1.7 &  1.5 &  1.3 & 0.565E-07 &  0.4207 $\pm$  0.0061 $\pm$  0.0110 \\
  0.0250 &   8.890 &   0.694 &  0.1 &  0.5 &  2.0 &  1.7 &  1.5 & 0.358E-07 &  0.4427 $\pm$  0.0061 $\pm$  0.0136 \\
  0.0250 &  10.800 &   0.833 &  0.1 &  0.5 &  2.1 &  2.1 &  1.9 & 0.261E-07 &  0.4732 $\pm$  0.0112 $\pm$  0.0169 \\
 \hline
  0.0350 &   1.300 &   0.229 &  1.3 &  0.7 &  0.3 &  0.2 &  0.1 & 0.124E-05 &  0.3183 $\pm$  0.0087 $\pm$  0.0048 \\
  0.0350 &   1.760 &   0.307 &  1.0 &  0.5 &  0.2 &  0.2 &  0.1 & 0.665E-06 &  0.3429 $\pm$  0.0061 $\pm$  0.0040 \\
  0.0350 &   2.450 &   0.396 &  0.8 &  0.2 &  0.4 &  0.3 &  0.2 & 0.330E-06 &  0.3716 $\pm$  0.0043 $\pm$  0.0038 \\
  0.0350 &   3.420 &   0.491 &  0.7 &  0.1 &  0.6 &  0.8 &  0.3 & 0.164E-06 &  0.3905 $\pm$  0.0060 $\pm$  0.0048 \\
  0.0350 &   4.390 &   0.584 &  0.6 &  0.1 &  0.8 &  1.5 &  0.5 & 0.963E-07 &  0.3939 $\pm$  0.0098 $\pm$  0.0074 \\
  0.0350 &   5.460 &   0.458 &  0.6 &  0.0 &  1.0 &  0.9 &  0.8 & 0.694E-07 &  0.3946 $\pm$  0.0128 $\pm$  0.0067 \\
  0.0350 &   6.960 &   0.542 &  0.6 &  0.1 &  1.2 &  1.1 &  1.1 & 0.422E-07 &  0.4257 $\pm$  0.0101 $\pm$  0.0089 \\
  0.0350 &   9.000 &   0.531 &  0.1 &  0.0 &  1.5 &  1.0 &  1.2 & 0.262E-07 &  0.4121 $\pm$  0.0074 $\pm$  0.0088 \\
  0.0350 &  11.440 &   0.638 &  0.1 &  0.5 &  1.7 &  1.6 &  1.4 & 0.157E-07 &  0.4355 $\pm$  0.0061 $\pm$  0.0122 \\
  0.0350 &  14.110 &   0.777 &  0.1 &  0.5 &  2.0 &  2.2 &  1.7 & 0.104E-07 &  0.4531 $\pm$  0.0104 $\pm$  0.0156 \\
 \hline
 \end{tabular}
 \end{table}
 \newpage
\begin{table}[t]                                                                                                                   
\begin{tabular}{|c|c|c||c|c|c|c|c||c|r|}\hline
$x$ & $Q^2$ & $y$ & $E$ & $E^{\prime}$ & $AC$ & $RC$ & $RE$& $d^2\sigma^{meas}/dxdQ^2$ &
$F^p_2\;\pm\Delta{F^{stat}_2}\;\pm\Delta{F^{syst}_2}$ \\
  & $[{\rm GeV^2}]$ & &  [\%]  &  [\%]  &  [\%]  &  [\%]  &  [\%]  & $[{\rm b\cdot GeV^{-2}}]$ & \\ \hline
 \hline
  0.0500 &   1.330 &   0.166 &  1.7 &  1.1 &  0.9 &  0.1 &  0.1 & 0.889E-06 &  0.3422 $\pm$  0.0101 $\pm$  0.0077 \\
  0.0500 &   1.760 &   0.218 &  1.3 &  0.8 &  0.4 &  0.1 &  0.1 & 0.504E-06 &  0.3413 $\pm$  0.0059 $\pm$  0.0054 \\
  0.0500 &   2.480 &   0.287 &  1.0 &  0.5 &  0.2 &  0.2 &  0.1 & 0.247E-06 &  0.3539 $\pm$  0.0037 $\pm$  0.0041 \\
  0.0500 &   3.440 &   0.367 &  0.8 &  0.3 &  0.2 &  0.4 &  0.2 & 0.123E-06 &  0.3841 $\pm$  0.0045 $\pm$  0.0039 \\
  0.0500 &   4.450 &   0.436 &  0.7 &  0.1 &  0.3 &  0.8 &  0.3 & 0.717E-07 &  0.3796 $\pm$  0.0063 $\pm$  0.0044 \\
  0.0500 &   5.440 &   0.417 &  0.7 &  0.0 &  0.5 &  0.7 &  0.5 & 0.478E-07 &  0.3836 $\pm$  0.0083 $\pm$  0.0046 \\
  0.0500 &   6.890 &   0.415 &  0.6 &  0.0 &  0.7 &  0.6 &  0.8 & 0.314E-07 &  0.3874 $\pm$  0.0079 $\pm$  0.0053 \\
  0.0500 &   9.060 &   0.422 &  0.1 &  0.1 &  0.9 &  0.7 &  0.9 & 0.187E-07 &  0.3957 $\pm$  0.0072 $\pm$  0.0059 \\
  0.0500 &  11.490 &   0.479 &  0.1 &  0.0 &  1.2 &  0.9 &  1.0 & 0.115E-07 &  0.4202 $\pm$  0.0054 $\pm$  0.0076 \\
  0.0500 &  14.870 &   0.583 &  0.1 &  0.5 &  1.4 &  1.4 &  1.3 & 0.657E-08 &  0.4229 $\pm$  0.0051 $\pm$  0.0103 \\
  0.0500 &  18.910 &   0.731 &  0.1 &  0.5 &  1.7 &  2.1 &  1.6 & 0.392E-08 &  0.4486 $\pm$  0.0077 $\pm$  0.0142 \\
 \hline
  0.0700 &   1.350 &   0.120 &  2.3 &  1.7 &  1.1 &  0.1 &  0.1 & 0.654E-06 &  0.3232 $\pm$  0.0142 $\pm$  0.0099 \\
  0.0700 &   1.770 &   0.156 &  1.8 &  1.2 &  0.6 &  0.1 &  0.1 & 0.378E-06 &  0.3536 $\pm$  0.0086 $\pm$  0.0078 \\
  0.0700 &   2.490 &   0.207 &  1.3 &  0.8 &  0.3 &  0.1 &  0.1 & 0.187E-06 &  0.3580 $\pm$  0.0050 $\pm$  0.0056 \\
  0.0700 &   3.470 &   0.270 &  1.1 &  0.5 &  0.3 &  0.2 &  0.1 & 0.936E-07 &  0.3829 $\pm$  0.0053 $\pm$  0.0046 \\
  0.0700 &   4.450 &   0.325 &  0.9 &  0.3 &  0.3 &  0.4 &  0.2 & 0.549E-07 &  0.3799 $\pm$  0.0067 $\pm$  0.0041 \\
  0.0700 &   5.450 &   0.376 &  0.8 &  0.2 &  0.3 &  0.5 &  0.3 & 0.356E-07 &  0.3869 $\pm$  0.0092 $\pm$  0.0040 \\
  0.0700 &   6.890 &   0.337 &  0.7 &  0.1 &  0.4 &  0.4 &  0.5 & 0.226E-07 &  0.3992 $\pm$  0.0097 $\pm$  0.0042 \\
  0.0700 &   8.920 &   0.366 &  0.7 &  0.0 &  0.6 &  0.5 &  0.8 & 0.138E-07 &  0.3997 $\pm$  0.0110 $\pm$  0.0051 \\
  0.0700 &  11.510 &   0.365 &  0.1 &  0.1 &  0.8 &  0.5 &  0.8 & 0.848E-08 &  0.4155 $\pm$  0.0075 $\pm$  0.0052 \\
  0.0700 &  14.920 &   0.434 &  0.1 &  0.0 &  1.1 &  0.7 &  0.9 & 0.488E-08 &  0.4034 $\pm$  0.0059 $\pm$  0.0065 \\
  0.0700 &  19.650 &   0.546 &  0.1 &  0.6 &  1.4 &  1.3 &  1.2 & 0.266E-08 &  0.4191 $\pm$  0.0060 $\pm$  0.0096 \\
  0.0700 &  25.560 &   0.705 &  0.1 &  0.5 &  1.7 &  2.2 &  1.5 & 0.148E-08 &  0.4630 $\pm$  0.0132 $\pm$  0.0151 \\
 \hline
  0.0900 &   1.380 &   0.096 &  2.9 &  2.3 &  0.8 &  0.0 &  0.1 & 0.508E-06 &  0.3456 $\pm$  0.0214 $\pm$  0.0130 \\
  0.0900 &   1.760 &   0.121 &  2.2 &  1.6 &  0.4 &  0.0 &  0.1 & 0.309E-06 &  0.3475 $\pm$  0.0110 $\pm$  0.0097 \\
  0.0900 &   2.490 &   0.163 &  1.6 &  1.0 &  0.2 &  0.1 &  0.1 & 0.151E-06 &  0.3721 $\pm$  0.0066 $\pm$  0.0073 \\
  0.0900 &   3.470 &   0.213 &  1.3 &  0.7 &  0.3 &  0.1 &  0.1 & 0.753E-07 &  0.3677 $\pm$  0.0061 $\pm$  0.0054 \\
  0.0900 &   4.450 &   0.257 &  1.1 &  0.4 &  0.4 &  0.2 &  0.1 & 0.444E-07 &  0.3756 $\pm$  0.0074 $\pm$  0.0046 \\
  0.0900 &   5.460 &   0.300 &  0.9 &  0.3 &  0.4 &  0.3 &  0.2 & 0.286E-07 &  0.3771 $\pm$  0.0096 $\pm$  0.0043 \\
  0.0900 &   6.810 &   0.293 &  0.9 &  0.2 &  0.5 &  0.3 &  0.4 & 0.183E-07 &  0.3899 $\pm$  0.0093 $\pm$  0.0044 \\
  0.0900 &   8.900 &   0.301 &  0.8 &  0.1 &  0.7 &  0.3 &  0.6 & 0.109E-07 &  0.3778 $\pm$  0.0119 $\pm$  0.0048 \\
  0.0900 &  11.560 &   0.290 &  0.1 &  0.3 &  1.0 &  0.3 &  0.7 & 0.668E-08 &  0.3925 $\pm$  0.0091 $\pm$  0.0050 \\
  0.0900 &  14.910 &   0.349 &  0.1 &  0.1 &  1.2 &  0.4 &  0.7 & 0.388E-08 &  0.4007 $\pm$  0.0074 $\pm$  0.0060 \\
  0.0900 &  19.820 &   0.436 &  0.1 &  0.0 &  1.5 &  0.7 &  0.9 & 0.209E-08 &  0.3949 $\pm$  0.0063 $\pm$  0.0077 \\
  0.0900 &  26.070 &   0.560 &  0.1 &  0.6 &  1.9 &  1.4 &  1.2 & 0.112E-08 &  0.3998 $\pm$  0.0082 $\pm$  0.0106 \\
 \hline
  0.1100 &   1.780 &   0.100 &  2.7 &  2.1 &  0.6 &  0.0 &  0.1 & 0.253E-06 &  0.3528 $\pm$  0.0135 $\pm$  0.0121 \\
  0.1100 &   2.500 &   0.134 &  1.9 &  1.3 &  0.2 &  0.0 &  0.1 & 0.124E-06 &  0.3620 $\pm$  0.0077 $\pm$  0.0085 \\
  0.1100 &   3.470 &   0.178 &  1.5 &  0.8 &  0.3 &  0.1 &  0.1 & 0.625E-07 &  0.3766 $\pm$  0.0075 $\pm$  0.0064 \\
  0.1100 &   4.460 &   0.216 &  1.2 &  0.6 &  0.4 &  0.1 &  0.1 & 0.368E-07 &  0.3735 $\pm$  0.0083 $\pm$  0.0052 \\
  0.1100 &   5.470 &   0.250 &  1.1 &  0.4 &  0.4 &  0.1 &  0.1 & 0.238E-07 &  0.3708 $\pm$  0.0101 $\pm$  0.0046 \\
  0.1100 &   6.820 &   0.251 &  1.0 &  0.3 &  0.6 &  0.2 &  0.3 & 0.152E-07 &  0.3662 $\pm$  0.0093 $\pm$  0.0044 \\
  0.1100 &   8.930 &   0.258 &  0.9 &  0.2 &  0.8 &  0.2 &  0.5 & 0.888E-08 &  0.3762 $\pm$  0.0129 $\pm$  0.0050 \\
  0.1100 &  11.500 &   0.248 &  0.2 &  0.1 &  1.1 &  0.2 &  0.7 & 0.549E-08 &  0.3662 $\pm$  0.0100 $\pm$  0.0047 \\
  0.1100 &  14.920 &   0.291 &  0.1 &  0.2 &  1.3 &  0.2 &  0.7 & 0.319E-08 &  0.3795 $\pm$  0.0085 $\pm$  0.0058 \\
  0.1100 &  19.800 &   0.360 &  0.1 &  0.0 &  1.7 &  0.4 &  0.8 & 0.174E-08 &  0.3850 $\pm$  0.0070 $\pm$  0.0072 \\
  0.1100 &  26.370 &   0.464 &  0.1 &  0.6 &  2.0 &  0.7 &  1.0 & 0.915E-09 &  0.3724 $\pm$  0.0077 $\pm$  0.0089 \\
  0.1100 &  34.480 &   0.604 &  0.2 &  0.5 &  2.3 &  1.4 &  1.3 & 0.492E-09 &  0.4285 $\pm$  0.0167 $\pm$  0.0130 \\
 \hline
 \end{tabular}
 \end{table}
 \newpage
\begin{table}[t]                                                                                                                   
\begin{tabular}{|c|c|c||c|c|c|c|c||c|r|}\hline
$x$ & $Q^2$ & $y$ & $E$ & $E^{\prime}$ & $AC$ & $RC$ & $RE$& $d^2\sigma^{meas}/dxdQ^2$ &
$F^p_2\;\pm\Delta{F^{stat}_2}\;\pm\Delta{F^{syst}_2}$ \\
  & $[{\rm GeV^2}]$ & &  [\%]  &  [\%]  &  [\%]  &  [\%]  &  [\%]  & $[{\rm b\cdot GeV^{-2}}]$ & \\ \hline
 \hline
  0.1400 &   1.860 &   0.082 &  3.2 &  2.6 &  1.1 &  0.0 &  0.1 & 0.188E-06 &  0.3772 $\pm$  0.0224 $\pm$  0.0162 \\
  0.1400 &   2.490 &   0.110 &  2.3 &  1.7 &  0.7 &  0.0 &  0.1 & 0.101E-06 &  0.3704 $\pm$  0.0073 $\pm$  0.0111 \\
  0.1400 &   3.480 &   0.141 &  1.7 &  1.1 &  0.3 &  0.0 &  0.1 & 0.496E-07 &  0.3743 $\pm$  0.0064 $\pm$  0.0078 \\
  0.1400 &   4.460 &   0.172 &  1.4 &  0.8 &  0.2 &  0.1 &  0.1 & 0.292E-07 &  0.3579 $\pm$  0.0065 $\pm$  0.0059 \\
  0.1400 &   5.470 &   0.201 &  1.2 &  0.6 &  0.3 &  0.1 &  0.1 & 0.190E-07 &  0.3431 $\pm$  0.0076 $\pm$  0.0048 \\
  0.1400 &   6.840 &   0.212 &  1.1 &  0.4 &  0.5 &  0.1 &  0.2 & 0.118E-07 &  0.3551 $\pm$  0.0070 $\pm$  0.0047 \\
  0.1400 &   8.940 &   0.213 &  1.0 &  0.3 &  0.8 &  0.1 &  0.4 & 0.686E-08 &  0.3742 $\pm$  0.0112 $\pm$  0.0053 \\
  0.1400 &  11.390 &   0.217 &  0.2 &  0.2 &  1.1 &  0.1 &  0.7 & 0.426E-08 &  0.3692 $\pm$  0.0091 $\pm$  0.0050 \\
  0.1400 &  14.960 &   0.233 &  0.2 &  0.3 &  1.4 &  0.1 &  0.6 & 0.247E-08 &  0.3646 $\pm$  0.0069 $\pm$  0.0058 \\
  0.1400 &  19.790 &   0.289 &  0.2 &  0.2 &  1.7 &  0.2 &  0.6 & 0.136E-08 &  0.3591 $\pm$  0.0054 $\pm$  0.0068 \\
  0.1400 &  26.550 &   0.374 &  0.2 &  0.0 &  2.1 &  0.4 &  0.8 & 0.711E-09 &  0.3683 $\pm$  0.0058 $\pm$  0.0084 \\
  0.1400 &  35.200 &   0.489 &  0.2 &  0.5 &  2.4 &  0.9 &  1.0 & 0.373E-09 &  0.3572 $\pm$  0.0098 $\pm$  0.0100 \\
 \hline
  0.1800 &   2.670 &   0.092 &  2.7 &  2.0 &  1.2 &  0.0 &  0.1 & 0.685E-07 &  0.3701 $\pm$  0.0111 $\pm$  0.0131 \\
  0.1800 &   3.480 &   0.114 &  2.1 &  1.4 &  0.7 &  0.0 &  0.1 & 0.384E-07 &  0.3397 $\pm$  0.0075 $\pm$  0.0088 \\
  0.1800 &   4.470 &   0.135 &  1.7 &  1.0 &  0.3 &  0.0 &  0.1 & 0.225E-07 &  0.3646 $\pm$  0.0080 $\pm$  0.0072 \\
  0.1800 &   5.470 &   0.158 &  1.4 &  0.7 &  0.1 &  0.0 &  0.1 & 0.146E-07 &  0.3304 $\pm$  0.0085 $\pm$  0.0054 \\
  0.1800 &   6.870 &   0.175 &  1.2 &  0.5 &  0.4 &  0.1 &  0.2 & 0.894E-08 &  0.3466 $\pm$  0.0078 $\pm$  0.0049 \\
  0.1800 &   8.890 &   0.174 &  1.1 &  0.4 &  0.8 &  0.1 &  0.4 & 0.525E-08 &  0.3446 $\pm$  0.0118 $\pm$  0.0049 \\
  0.1800 &  11.330 &   0.192 &  1.0 &  0.3 &  1.1 &  0.1 &  0.5 & 0.320E-08 &  0.3488 $\pm$  0.0109 $\pm$  0.0056 \\
  0.1800 &  14.970 &   0.183 &  0.2 &  0.4 &  1.4 &  0.1 &  0.6 & 0.186E-08 &  0.3333 $\pm$  0.0079 $\pm$  0.0054 \\
  0.1800 &  19.860 &   0.229 &  0.2 &  0.2 &  1.8 &  0.1 &  0.6 & 0.102E-08 &  0.3343 $\pm$  0.0060 $\pm$  0.0063 \\
  0.1800 &  26.660 &   0.294 &  0.2 &  0.1 &  2.1 &  0.3 &  0.6 & 0.536E-09 &  0.3359 $\pm$  0.0061 $\pm$  0.0074 \\
  0.1800 &  35.140 &   0.378 &  0.2 &  0.5 &  2.4 &  0.5 &  0.8 & 0.289E-09 &  0.3491 $\pm$  0.0081 $\pm$  0.0092 \\
  0.1800 &  45.750 &   0.490 &  0.2 &  0.5 &  2.7 &  0.9 &  1.0 & 0.157E-09 &  0.3650 $\pm$  0.0153 $\pm$  0.0111 \\
 \hline
  0.2250 &   2.900 &   0.080 &  2.8 &  2.1 &  1.0 &  0.0 &  0.1 & 0.456E-07 &  0.3389 $\pm$  0.0321 $\pm$  0.0123 \\
  0.2250 &   3.510 &   0.097 &  2.2 &  1.5 &  0.7 &  0.0 &  0.1 & 0.295E-07 &  0.3372 $\pm$  0.0089 $\pm$  0.0094 \\
  0.2250 &   4.490 &   0.112 &  1.8 &  1.1 &  0.3 &  0.0 &  0.1 & 0.172E-07 &  0.3258 $\pm$  0.0079 $\pm$  0.0069 \\
  0.2250 &   5.470 &   0.129 &  1.5 &  0.8 &  0.2 &  0.0 &  0.1 & 0.113E-07 &  0.3340 $\pm$  0.0090 $\pm$  0.0059 \\
  0.2250 &   6.880 &   0.154 &  1.3 &  0.6 &  0.5 &  0.0 &  0.1 & 0.684E-08 &  0.3234 $\pm$  0.0079 $\pm$  0.0049 \\
  0.2250 &   8.870 &   0.154 &  1.1 &  0.4 &  0.8 &  0.1 &  0.3 & 0.400E-08 &  0.3109 $\pm$  0.0103 $\pm$  0.0046 \\
  0.2250 &  11.370 &   0.158 &  1.0 &  0.3 &  1.1 &  0.1 &  0.5 & 0.240E-08 &  0.3445 $\pm$  0.0111 $\pm$  0.0056 \\
  0.2250 &  15.030 &   0.148 &  0.2 &  0.4 &  1.4 &  0.1 &  0.6 & 0.138E-08 &  0.3104 $\pm$  0.0080 $\pm$  0.0050 \\
  0.2250 &  19.860 &   0.185 &  0.2 &  0.2 &  1.7 &  0.1 &  0.6 & 0.764E-09 &  0.3034 $\pm$  0.0058 $\pm$  0.0056 \\
  0.2250 &  26.720 &   0.237 &  0.2 &  0.1 &  2.0 &  0.2 &  0.6 & 0.403E-09 &  0.3117 $\pm$  0.0060 $\pm$  0.0066 \\
  0.2250 &  35.290 &   0.305 &  0.2 &  0.5 &  2.3 &  0.3 &  0.6 & 0.218E-09 &  0.2986 $\pm$  0.0070 $\pm$  0.0073 \\
  0.2250 &  46.570 &   0.401 &  0.2 &  0.5 &  2.5 &  0.6 &  0.8 & 0.116E-09 &  0.3114 $\pm$  0.0118 $\pm$  0.0087 \\
 \hline
  0.2750 &   3.750 &   0.085 &  2.2 &  1.4 &  1.6 &  0.0 &  0.1 & 0.201E-07 &  0.3063 $\pm$  0.0149 $\pm$  0.0093 \\
  0.2750 &   4.470 &   0.101 &  1.8 &  1.0 &  1.1 &  0.0 &  0.1 & 0.134E-07 &  0.3105 $\pm$  0.0104 $\pm$  0.0072 \\
  0.2750 &   5.480 &   0.109 &  1.5 &  0.7 &  0.8 &  0.0 &  0.1 & 0.859E-08 &  0.3058 $\pm$  0.0101 $\pm$  0.0057 \\
  0.2750 &   6.890 &   0.128 &  1.2 &  0.5 &  0.9 &  0.0 &  0.1 & 0.519E-08 &  0.3038 $\pm$  0.0086 $\pm$  0.0050 \\
  0.2750 &   8.820 &   0.156 &  1.0 &  0.3 &  1.1 &  0.0 &  0.1 & 0.302E-08 &  0.3032 $\pm$  0.0125 $\pm$  0.0047 \\
  0.2750 &  11.450 &   0.135 &  1.0 &  0.2 &  1.3 &  0.0 &  0.4 & 0.177E-08 &  0.2761 $\pm$  0.0111 $\pm$  0.0047 \\
  0.2750 &  14.910 &   0.143 &  0.9 &  0.0 &  1.5 &  0.1 &  0.5 & 0.102E-08 &  0.2900 $\pm$  0.0110 $\pm$  0.0053 \\
  0.2750 &  19.880 &   0.152 &  0.2 &  0.1 &  1.7 &  0.1 &  0.6 & 0.566E-09 &  0.2789 $\pm$  0.0065 $\pm$  0.0051 \\
  0.2750 &  26.740 &   0.195 &  0.2 &  0.1 &  1.9 &  0.1 &  0.5 & 0.299E-09 &  0.2610 $\pm$  0.0060 $\pm$  0.0052 \\
  0.2750 &  35.340 &   0.253 &  0.2 &  0.2 &  2.1 &  0.2 &  0.5 & 0.162E-09 &  0.2778 $\pm$  0.0071 $\pm$  0.0061 \\
  0.2750 &  46.990 &   0.332 &  0.3 &  0.5 &  2.3 &  0.4 &  0.6 & 0.856E-10 &  0.2726 $\pm$  0.0110 $\pm$  0.0067 \\
  0.2750 &  59.790 &   0.418 &  0.3 &  0.5 &  2.4 &  0.7 &  1.3 & 0.492E-10 &  0.1865 $\pm$  0.0284 $\pm$  0.0054 \\
 \hline
 \end{tabular}
 \end{table}
 \newpage
\begin{table}[t]                                                                                                                   
\begin{tabular}{|c|c|c||c|c|c|c|c||c|r|}\hline
$x$ & $Q^2$ & $y$ & $E$ & $E^{\prime}$ & $AC$ & $RC$ & $RE$& $d^2\sigma^{meas}/dxdQ^2$ &
$F^p_2\;\pm\Delta{F^{stat}_2}\;\pm\Delta{F^{syst}_2}$ \\
  & $[{\rm GeV^2}]$ & &  [\%]  &  [\%]  &  [\%]  &  [\%]  &  [\%]  & $[{\rm b\cdot GeV^{-2}}]$ & \\ \hline
 \hline
  0.3500 &   4.620 &   0.082 &  1.5 &  0.7 &  4.6 &  0.0 &  0.1 & 0.882E-08 &  0.2773 $\pm$  0.0118 $\pm$  0.0136 \\
  0.3500 &   5.460 &   0.097 &  1.2 &  0.3 &  3.6 &  0.0 &  0.1 & 0.589E-08 &  0.2578 $\pm$  0.0095 $\pm$  0.0098 \\
  0.3500 &   6.970 &   0.106 &  0.9 &  0.0 &  2.3 &  0.0 &  0.1 & 0.341E-08 &  0.2685 $\pm$  0.0072 $\pm$  0.0066 \\
  0.3500 &   8.820 &   0.124 &  0.7 &  0.2 &  2.1 &  0.0 &  0.1 & 0.202E-08 &  0.2482 $\pm$  0.0087 $\pm$  0.0056 \\
  0.3500 &  11.320 &   0.150 &  0.6 &  0.3 &  2.0 &  0.1 &  0.2 & 0.116E-08 &  0.2331 $\pm$  0.0117 $\pm$  0.0049 \\
  0.3500 &  14.830 &   0.126 &  0.5 &  0.4 &  1.9 &  0.0 &  0.6 & 0.663E-09 &  0.2171 $\pm$  0.0080 $\pm$  0.0045 \\
  0.3500 &  20.240 &   0.128 &  0.4 &  0.5 &  1.8 &  0.0 &  0.6 & 0.348E-09 &  0.2229 $\pm$  0.0057 $\pm$  0.0045 \\
  0.3500 &  26.670 &   0.155 &  0.3 &  0.5 &  2.0 &  0.1 &  0.5 & 0.193E-09 &  0.2131 $\pm$  0.0043 $\pm$  0.0045 \\
  0.3500 &  35.420 &   0.201 &  0.3 &  0.5 &  2.3 &  0.1 &  0.5 & 0.104E-09 &  0.2135 $\pm$  0.0048 $\pm$  0.0053 \\
  0.3500 &  46.650 &   0.259 &  0.3 &  0.5 &  2.6 &  0.2 &  0.5 & 0.565E-10 &  0.2180 $\pm$  0.0064 $\pm$  0.0060 \\
  0.3500 &  61.210 &   0.337 &  0.4 &  0.5 &  2.8 &  0.4 &  1.2 & 0.305E-10 &  0.2312 $\pm$  0.0190 $\pm$  0.0073 \\
 \hline
  0.5000 &   5.730 &   0.072 &  1.4 &  2.6 & 11.0 &  0.0 &  0.1 & 0.252E-08 &  0.1569 $\pm$  0.0158 $\pm$  0.0178 \\
  0.5000 &   6.930 &   0.087 &  1.3 &  2.5 &  9.1 &  0.0 &  0.1 & 0.156E-08 &  0.1680 $\pm$  0.0075 $\pm$  0.0160 \\
  0.5000 &   8.880 &   0.093 &  1.4 &  2.6 &  6.8 &  0.0 &  0.1 & 0.848E-09 &  0.1356 $\pm$  0.0062 $\pm$  0.0100 \\
  0.5000 &  11.270 &   0.105 &  1.5 &  2.7 &  4.7 &  0.0 &  0.2 & 0.481E-09 &  0.1412 $\pm$  0.0074 $\pm$  0.0079 \\
  0.5000 &  14.370 &   0.133 &  1.2 &  2.4 &  2.7 &  0.0 &  0.2 & 0.274E-09 &  0.1321 $\pm$  0.0099 $\pm$  0.0051 \\
  0.5000 &  20.050 &   0.110 &  2.0 &  3.2 &  1.5 &  0.0 &  0.7 & 0.133E-09 &  0.1167 $\pm$  0.0057 $\pm$  0.0048 \\
  0.5000 &  27.240 &   0.115 &  0.7 &  0.6 &  1.8 &  0.0 &  0.5 & 0.677E-10 &  0.1154 $\pm$  0.0034 $\pm$  0.0025 \\
  0.5000 &  35.510 &   0.142 &  0.7 &  0.6 &  3.0 &  0.1 &  0.5 & 0.378E-10 &  0.1086 $\pm$  0.0027 $\pm$  0.0034 \\
  0.5000 &  46.630 &   0.183 &  0.7 &  0.5 &  3.9 &  0.1 &  0.4 & 0.206E-10 &  0.1012 $\pm$  0.0033 $\pm$  0.0041 \\
  0.5000 &  62.340 &   0.243 &  0.7 &  0.5 &  4.6 &  0.3 &  1.1 & 0.107E-10 &  0.1081 $\pm$  0.0086 $\pm$  0.0052 \\
 \hline
 \end{tabular}
 \end{table}

\begin{table}[t]                                                                                                                   

~
\vspace{1cm}
\caption{
The deuteron structure function $F_2^d(x,Q^2)$ from the measurements at 90, 120,
200 and 280~GeV, averaged over all energies.
In this table the value of $F_2$ and its statistical and
total systematic errors are given for each $x$ and $Q^2$.
The mean $y$ of the events in this bin is also given.
Also listed is the measured cross section (${\rm d}^2\sigma^{meas}/
{\rm d}x{\rm d}Q^2$) at the same $x$ and $Q^2$.
The systematic error is the quadratic sum of the contributions given as
percentages of $F_2$ in the columns 4--8.  The contributions are:
E, E', errors due to calibrations of
incident and scattered muon energies;
AC, error in the determination of the spectrometer acceptance;
RC, error from the radiative corrections;
RE, error from the reconstruction inefficiency due to showers.
}
\label{tab:d2}
\vspace{1cm}

\begin{tabular}{|c|c|c||c|c|c|c|c||c|r|}\hline
$x$ & $Q^2$ & $y$ & $E$ & $E^{\prime}$ & $AC$ & $RC$ & $RE$& $d^2\sigma^{meas}/dxdQ^2$ &
$F^d_2\;\pm\Delta{F^{stat}_2}\;\pm\Delta{F^{syst}_2}$ \\
  & $[{\rm GeV^2}]$ & &  [\%]  &  [\%]  &  [\%]  &  [\%]  &  [\%]  & $[{\rm b\cdot GeV^{-2}}]$ & \\ 
 \hline
  0.0080 &   0.800 &   0.617 &  0.6 &  0.1 &  2.1 &  1.7 &  0.3 & 0.874E-05 &  0.2698 $\pm$  0.0026 $\pm$  0.0075 \\
  0.0080 &   1.120 &   0.747 &  0.5 &  0.0 &  1.0 &  1.9 &  0.5 & 0.452E-05 &  0.2890 $\pm$  0.0044 $\pm$  0.0064 \\
  0.0080 &   3.460 &   0.854 &  0.1 &  0.5 &  1.8 &  1.9 &  2.0 & 0.670E-06 &  0.3755 $\pm$  0.0086 $\pm$  0.0124 \\
 \hline
  0.0125 &   0.900 &   0.444 &  0.7 &  0.3 &  2.7 &  0.8 &  0.2 & 0.523E-05 &  0.2769 $\pm$  0.0054 $\pm$  0.0080 \\
  0.0125 &   1.230 &   0.582 &  0.6 &  0.1 &  1.7 &  1.3 &  0.3 & 0.266E-05 &  0.3015 $\pm$  0.0027 $\pm$  0.0068 \\
  0.0125 &   1.670 &   0.667 &  0.5 &  0.0 &  1.0 &  1.4 &  0.5 & 0.149E-05 &  0.3230 $\pm$  0.0044 $\pm$  0.0060 \\
  0.0125 &   2.730 &   0.604 &  0.6 &  0.1 &  1.0 &  1.1 &  1.2 & 0.644E-06 &  0.3597 $\pm$  0.0143 $\pm$  0.0071 \\
  0.0125 &   3.250 &   0.709 &  0.5 &  0.0 &  1.1 &  1.3 &  1.3 & 0.445E-06 &  0.3583 $\pm$  0.0140 $\pm$  0.0078 \\
  0.0125 &   4.520 &   0.713 &  0.1 &  0.5 &  1.2 &  2.4 &  1.7 & 0.245E-06 &  0.3829 $\pm$  0.0060 $\pm$  0.0122 \\
  0.0125 &   5.380 &   0.838 &  0.1 &  0.5 &  1.3 &  2.6 &  1.9 & 0.183E-06 &  0.4169 $\pm$  0.0078 $\pm$  0.0148 \\
 \hline
  0.0175 &   1.270 &   0.444 &  0.7 &  0.2 &  2.0 &  0.7 &  0.2 & 0.204E-05 &  0.3082 $\pm$  0.0041 $\pm$  0.0070 \\
  0.0175 &   1.730 &   0.559 &  0.6 &  0.1 &  1.3 &  0.9 &  0.3 & 0.106E-05 &  0.3411 $\pm$  0.0037 $\pm$  0.0061 \\
  0.0175 &   2.280 &   0.637 &  0.5 &  0.0 &  1.0 &  1.1 &  0.5 & 0.606E-06 &  0.3436 $\pm$  0.0045 $\pm$  0.0055 \\
  0.0175 &   3.620 &   0.567 &  0.6 &  0.1 &  1.1 &  1.2 &  1.1 & 0.277E-06 &  0.3737 $\pm$  0.0104 $\pm$  0.0076 \\
  0.0175 &   4.330 &   0.673 &  0.5 &  0.0 &  1.2 &  1.8 &  1.3 & 0.187E-06 &  0.3945 $\pm$  0.0122 $\pm$  0.0100 \\
  0.0175 &   5.600 &   0.631 &  0.1 &  0.5 &  1.3 &  2.0 &  1.5 & 0.120E-06 &  0.3974 $\pm$  0.0076 $\pm$  0.0113 \\
  0.0175 &   6.730 &   0.748 &  0.1 &  0.5 &  1.4 &  2.2 &  1.7 & 0.821E-07 &  0.4130 $\pm$  0.0052 $\pm$  0.0131 \\
 \hline
  0.0250 &   1.270 &   0.315 &  1.0 &  0.5 &  2.2 &  0.3 &  0.1 & 0.160E-05 &  0.3063 $\pm$  0.0050 $\pm$  0.0075 \\
  0.0250 &   1.760 &   0.416 &  0.8 &  0.3 &  1.4 &  0.5 &  0.2 & 0.805E-06 &  0.3325 $\pm$  0.0034 $\pm$  0.0056 \\
  0.0250 &   2.410 &   0.519 &  0.6 &  0.1 &  0.9 &  0.7 &  0.3 & 0.412E-06 &  0.3513 $\pm$  0.0027 $\pm$  0.0048 \\
  0.0250 &   3.340 &   0.629 &  0.6 &  0.0 &  0.9 &  1.2 &  0.5 & 0.210E-06 &  0.3650 $\pm$  0.0052 $\pm$  0.0060 \\
  0.0250 &   4.540 &   0.500 &  0.7 &  0.1 &  1.0 &  1.1 &  1.0 & 0.131E-06 &  0.3871 $\pm$  0.0090 $\pm$  0.0075 \\
  0.0250 &   5.450 &   0.595 &  0.6 &  0.0 &  1.2 &  1.5 &  1.1 & 0.871E-07 &  0.3906 $\pm$  0.0085 $\pm$  0.0090 \\
  0.0250 &   7.090 &   0.586 &  0.1 &  0.1 &  1.4 &  1.4 &  1.3 & 0.537E-07 &  0.4011 $\pm$  0.0046 $\pm$  0.0095 \\
  0.0250 &   8.880 &   0.694 &  0.1 &  0.5 &  1.5 &  1.7 &  1.6 & 0.337E-07 &  0.4294 $\pm$  0.0046 $\pm$  0.0122 \\
  0.0250 &  10.790 &   0.832 &  0.1 &  0.5 &  1.7 &  2.1 &  1.9 & 0.238E-07 &  0.4460 $\pm$  0.0081 $\pm$  0.0149 \\
 \hline
  0.0350 &   1.300 &   0.229 &  1.3 &  0.7 &  2.2 &  0.4 &  0.1 & 0.119E-05 &  0.3141 $\pm$  0.0078 $\pm$  0.0085 \\
  0.0350 &   1.760 &   0.307 &  1.0 &  0.5 &  1.3 &  0.3 &  0.1 & 0.637E-06 &  0.3254 $\pm$  0.0050 $\pm$  0.0056 \\
  0.0350 &   2.450 &   0.398 &  0.8 &  0.2 &  0.7 &  0.4 &  0.2 & 0.314E-06 &  0.3537 $\pm$  0.0033 $\pm$  0.0042 \\
  0.0350 &   3.410 &   0.494 &  0.7 &  0.1 &  0.5 &  0.8 &  0.3 & 0.156E-06 &  0.3696 $\pm$  0.0043 $\pm$  0.0044 \\
  0.0350 &   4.390 &   0.585 &  0.6 &  0.0 &  0.7 &  1.4 &  0.5 & 0.908E-07 &  0.3704 $\pm$  0.0069 $\pm$  0.0065 \\
  0.0350 &   5.450 &   0.466 &  0.6 &  0.0 &  0.8 &  0.9 &  0.8 & 0.654E-07 &  0.3764 $\pm$  0.0097 $\pm$  0.0059 \\
  0.0350 &   6.950 &   0.541 &  0.6 &  0.1 &  1.1 &  1.1 &  1.1 & 0.400E-07 &  0.3925 $\pm$  0.0074 $\pm$  0.0077 \\
  0.0350 &   8.990 &   0.532 &  0.1 &  0.1 &  1.3 &  1.0 &  1.2 & 0.247E-07 &  0.3967 $\pm$  0.0056 $\pm$  0.0081 \\
  0.0350 &  11.440 &   0.637 &  0.1 &  0.5 &  1.6 &  1.5 &  1.4 & 0.148E-07 &  0.4163 $\pm$  0.0045 $\pm$  0.0111 \\
  0.0350 &  14.100 &   0.776 &  0.1 &  0.5 &  1.8 &  2.2 &  1.7 & 0.959E-08 &  0.4303 $\pm$  0.0074 $\pm$  0.0144 \\
 \hline
 \end{tabular}
 \end{table}
 \newpage
\begin{table}[t]                                                                                                                   
\begin{tabular}{|c|c|c||c|c|c|c|c||c|r|}\hline
$x$ & $Q^2$ & $y$ & $E$ & $E^{\prime}$ & $AC$ & $RC$ & $RE$& $d^2\sigma^{meas}/dxdQ^2$ &
$F^d_2\;\pm\Delta{F^{stat}_2}\;\pm\Delta{F^{syst}_2}$ \\
  & $[{\rm GeV^2}]$ & &  [\%]  &  [\%]  &  [\%]  &  [\%]  &  [\%]  & $[{\rm b\cdot GeV^{-2}}]$ & \\ \hline
 \hline
  0.0500 &   1.330 &   0.166 &  1.7 &  1.1 &  2.6 &  0.4 &  0.1 & 0.848E-06 &  0.3305 $\pm$  0.0091 $\pm$  0.0108 \\
  0.0500 &   1.760 &   0.218 &  1.3 &  0.8 &  1.3 &  0.4 &  0.1 & 0.480E-06 &  0.3266 $\pm$  0.0051 $\pm$  0.0067 \\
  0.0500 &   2.480 &   0.288 &  1.0 &  0.4 &  0.5 &  0.3 &  0.1 & 0.235E-06 &  0.3421 $\pm$  0.0030 $\pm$  0.0043 \\
  0.0500 &   3.440 &   0.370 &  0.8 &  0.2 &  0.2 &  0.4 &  0.2 & 0.117E-06 &  0.3636 $\pm$  0.0034 $\pm$  0.0036 \\
  0.0500 &   4.440 &   0.438 &  0.7 &  0.1 &  0.3 &  0.7 &  0.3 & 0.675E-07 &  0.3638 $\pm$  0.0046 $\pm$  0.0040 \\
  0.0500 &   5.430 &   0.435 &  0.7 &  0.0 &  0.4 &  0.7 &  0.5 & 0.449E-07 &  0.3759 $\pm$  0.0062 $\pm$  0.0043 \\
  0.0500 &   6.870 &   0.420 &  0.6 &  0.0 &  0.6 &  0.6 &  0.7 & 0.295E-07 &  0.3819 $\pm$  0.0061 $\pm$  0.0050 \\
  0.0500 &   9.050 &   0.425 &  0.1 &  0.1 &  0.9 &  0.6 &  0.9 & 0.176E-07 &  0.3817 $\pm$  0.0054 $\pm$  0.0056 \\
  0.0500 &  11.480 &   0.480 &  0.1 &  0.0 &  1.2 &  0.8 &  1.0 & 0.108E-07 &  0.3959 $\pm$  0.0039 $\pm$  0.0072 \\
  0.0500 &  14.870 &   0.583 &  0.1 &  0.5 &  1.5 &  1.3 &  1.3 & 0.614E-08 &  0.4029 $\pm$  0.0037 $\pm$  0.0099 \\
  0.0500 &  18.900 &   0.731 &  0.1 &  0.5 &  1.9 &  2.0 &  1.6 & 0.363E-08 &  0.4158 $\pm$  0.0053 $\pm$  0.0134 \\
 \hline
  0.0700 &   1.350 &   0.120 &  2.2 &  1.6 &  3.5 &  0.5 &  0.1 & 0.617E-06 &  0.3215 $\pm$  0.0134 $\pm$  0.0143 \\
  0.0700 &   1.770 &   0.156 &  1.7 &  1.1 &  1.8 &  0.4 &  0.1 & 0.357E-06 &  0.3297 $\pm$  0.0074 $\pm$  0.0092 \\
  0.0700 &   2.490 &   0.208 &  1.3 &  0.7 &  0.5 &  0.4 &  0.1 & 0.177E-06 &  0.3411 $\pm$  0.0042 $\pm$  0.0056 \\
  0.0700 &   3.460 &   0.271 &  1.0 &  0.4 &  0.2 &  0.3 &  0.1 & 0.880E-07 &  0.3602 $\pm$  0.0042 $\pm$  0.0042 \\
  0.0700 &   4.450 &   0.328 &  0.9 &  0.3 &  0.4 &  0.3 &  0.2 & 0.514E-07 &  0.3555 $\pm$  0.0049 $\pm$  0.0038 \\
  0.0700 &   5.440 &   0.379 &  0.8 &  0.2 &  0.5 &  0.5 &  0.3 & 0.334E-07 &  0.3649 $\pm$  0.0066 $\pm$  0.0040 \\
  0.0700 &   6.860 &   0.351 &  0.7 &  0.1 &  0.5 &  0.4 &  0.5 & 0.210E-07 &  0.3706 $\pm$  0.0070 $\pm$  0.0040 \\
  0.0700 &   8.910 &   0.369 &  0.7 &  0.0 &  0.5 &  0.4 &  0.7 & 0.129E-07 &  0.3655 $\pm$  0.0081 $\pm$  0.0045 \\
  0.0700 &  11.500 &   0.368 &  0.1 &  0.1 &  0.8 &  0.5 &  0.8 & 0.789E-08 &  0.3808 $\pm$  0.0055 $\pm$  0.0048 \\
  0.0700 &  14.900 &   0.437 &  0.1 &  0.0 &  1.1 &  0.7 &  1.0 & 0.454E-08 &  0.3866 $\pm$  0.0043 $\pm$  0.0064 \\
  0.0700 &  19.640 &   0.546 &  0.1 &  0.6 &  1.5 &  1.2 &  1.2 & 0.247E-08 &  0.3826 $\pm$  0.0041 $\pm$  0.0089 \\
  0.0700 &  25.560 &   0.705 &  0.1 &  0.5 &  1.9 &  2.2 &  1.5 & 0.136E-08 &  0.4119 $\pm$  0.0087 $\pm$  0.0137 \\
 \hline
  0.0900 &   1.380 &   0.096 &  2.7 &  2.2 &  4.4 &  0.4 &  0.1 & 0.474E-06 &  0.3322 $\pm$  0.0198 $\pm$  0.0187 \\
  0.0900 &   1.760 &   0.121 &  2.1 &  1.5 &  2.6 &  0.4 &  0.1 & 0.288E-06 &  0.3282 $\pm$  0.0097 $\pm$  0.0122 \\
  0.0900 &   2.480 &   0.163 &  1.6 &  1.0 &  0.9 &  0.4 &  0.1 & 0.141E-06 &  0.3445 $\pm$  0.0055 $\pm$  0.0073 \\
  0.0900 &   3.470 &   0.214 &  1.2 &  0.6 &  0.2 &  0.3 &  0.1 & 0.702E-07 &  0.3423 $\pm$  0.0049 $\pm$  0.0048 \\
  0.0900 &   4.450 &   0.259 &  1.0 &  0.4 &  0.5 &  0.3 &  0.1 & 0.412E-07 &  0.3584 $\pm$  0.0057 $\pm$  0.0045 \\
  0.0900 &   5.460 &   0.304 &  0.9 &  0.3 &  0.7 &  0.2 &  0.2 & 0.266E-07 &  0.3645 $\pm$  0.0071 $\pm$  0.0044 \\
  0.0900 &   6.800 &   0.305 &  0.8 &  0.2 &  0.8 &  0.4 &  0.3 & 0.170E-07 &  0.3610 $\pm$  0.0067 $\pm$  0.0045 \\
  0.0900 &   8.880 &   0.306 &  0.8 &  0.1 &  0.8 &  0.3 &  0.6 & 0.100E-07 &  0.3733 $\pm$  0.0092 $\pm$  0.0048 \\
  0.0900 &  11.550 &   0.294 &  0.2 &  0.2 &  0.8 &  0.3 &  0.7 & 0.618E-08 &  0.3618 $\pm$  0.0068 $\pm$  0.0041 \\
  0.0900 &  14.910 &   0.351 &  0.1 &  0.1 &  0.8 &  0.4 &  0.8 & 0.357E-08 &  0.3665 $\pm$  0.0053 $\pm$  0.0044 \\
  0.0900 &  19.790 &   0.437 &  0.1 &  0.1 &  1.2 &  0.7 &  0.9 & 0.193E-08 &  0.3665 $\pm$  0.0044 $\pm$  0.0062 \\
  0.0900 &  26.060 &   0.560 &  0.1 &  0.6 &  1.6 &  1.3 &  1.2 & 0.103E-08 &  0.3777 $\pm$  0.0056 $\pm$  0.0092 \\
 \hline
  0.1100 &   1.780 &   0.100 &  2.5 &  1.9 &  3.2 &  0.4 &  0.1 & 0.234E-06 &  0.3274 $\pm$  0.0119 $\pm$  0.0149 \\
  0.1100 &   2.500 &   0.134 &  1.8 &  1.2 &  1.4 &  0.4 &  0.1 & 0.114E-06 &  0.3363 $\pm$  0.0066 $\pm$  0.0089 \\
  0.1100 &   3.470 &   0.179 &  1.4 &  0.8 &  0.3 &  0.3 &  0.1 & 0.576E-07 &  0.3444 $\pm$  0.0060 $\pm$  0.0057 \\
  0.1100 &   4.460 &   0.218 &  1.2 &  0.5 &  0.4 &  0.3 &  0.1 & 0.338E-07 &  0.3422 $\pm$  0.0063 $\pm$  0.0047 \\
  0.1100 &   5.470 &   0.252 &  1.0 &  0.4 &  0.7 &  0.2 &  0.1 & 0.219E-07 &  0.3413 $\pm$  0.0074 $\pm$  0.0045 \\
  0.1100 &   6.800 &   0.259 &  0.9 &  0.3 &  0.9 &  0.3 &  0.3 & 0.139E-07 &  0.3445 $\pm$  0.0069 $\pm$  0.0046 \\
  0.1100 &   8.930 &   0.262 &  0.8 &  0.2 &  1.0 &  0.3 &  0.5 & 0.806E-08 &  0.3386 $\pm$  0.0093 $\pm$  0.0049 \\
  0.1100 &  11.460 &   0.251 &  0.2 &  0.1 &  1.0 &  0.2 &  0.7 & 0.503E-08 &  0.3442 $\pm$  0.0076 $\pm$  0.0044 \\
  0.1100 &  14.910 &   0.294 &  0.2 &  0.2 &  1.0 &  0.3 &  0.7 & 0.291E-08 &  0.3578 $\pm$  0.0064 $\pm$  0.0046 \\
  0.1100 &  19.800 &   0.360 &  0.2 &  0.0 &  1.0 &  0.4 &  0.8 & 0.158E-08 &  0.3558 $\pm$  0.0050 $\pm$  0.0047 \\
  0.1100 &  26.360 &   0.464 &  0.2 &  0.6 &  1.3 &  0.7 &  1.0 & 0.832E-09 &  0.3525 $\pm$  0.0054 $\pm$  0.0066 \\
  0.1100 &  34.470 &   0.603 &  0.2 &  0.5 &  1.7 &  1.4 &  1.3 & 0.446E-09 &  0.3801 $\pm$  0.0108 $\pm$  0.0098 \\
 \hline
 \end{tabular}
 \end{table}
 \newpage
\begin{table}[t]                                                                                                                   
\begin{tabular}{|c|c|c||c|c|c|c|c||c|r|}\hline
$x$ & $Q^2$ & $y$ & $E$ & $E^{\prime}$ & $AC$ & $RC$ & $RE$& $d^2\sigma^{meas}/dxdQ^2$ &
$F^d_2\;\pm\Delta{F^{stat}_2}\;\pm\Delta{F^{syst}_2}$ \\
  & $[{\rm GeV^2}]$ & &  [\%]  &  [\%]  &  [\%]  &  [\%]  &  [\%]  & $[{\rm b\cdot GeV^{-2}}]$ & \\ \hline
 \hline
  0.1400 &   1.860 &   0.082 &  3.0 &  2.4 &  3.5 &  0.4 &  0.1 & 0.171E-06 &  0.3536 $\pm$  0.0201 $\pm$  0.0185 \\
  0.1400 &   2.480 &   0.110 &  2.2 &  1.6 &  2.0 &  0.4 &  0.1 & 0.922E-07 &  0.3345 $\pm$  0.0061 $\pm$  0.0113 \\
  0.1400 &   3.480 &   0.142 &  1.6 &  1.0 &  0.7 &  0.3 &  0.1 & 0.450E-07 &  0.3337 $\pm$  0.0051 $\pm$  0.0070 \\
  0.1400 &   4.460 &   0.173 &  1.4 &  0.7 &  0.3 &  0.3 &  0.1 & 0.265E-07 &  0.3236 $\pm$  0.0050 $\pm$  0.0051 \\
  0.1400 &   5.470 &   0.202 &  1.2 &  0.5 &  0.5 &  0.3 &  0.1 & 0.171E-07 &  0.3301 $\pm$  0.0059 $\pm$  0.0046 \\
  0.1400 &   6.830 &   0.218 &  1.0 &  0.4 &  0.8 &  0.2 &  0.2 & 0.107E-07 &  0.3245 $\pm$  0.0051 $\pm$  0.0044 \\
  0.1400 &   8.930 &   0.221 &  0.9 &  0.3 &  1.0 &  0.2 &  0.4 & 0.617E-08 &  0.3368 $\pm$  0.0081 $\pm$  0.0050 \\
  0.1400 &  11.370 &   0.220 &  0.3 &  0.2 &  1.2 &  0.3 &  0.7 & 0.383E-08 &  0.3224 $\pm$  0.0065 $\pm$  0.0046 \\
  0.1400 &  14.940 &   0.235 &  0.2 &  0.2 &  1.3 &  0.2 &  0.7 & 0.222E-08 &  0.3269 $\pm$  0.0051 $\pm$  0.0049 \\
  0.1400 &  19.770 &   0.290 &  0.2 &  0.1 &  1.4 &  0.2 &  0.7 & 0.122E-08 &  0.3236 $\pm$  0.0038 $\pm$  0.0050 \\
  0.1400 &  26.540 &   0.374 &  0.2 &  0.1 &  1.4 &  0.4 &  0.8 & 0.636E-09 &  0.3247 $\pm$  0.0039 $\pm$  0.0055 \\
  0.1400 &  35.200 &   0.488 &  0.2 &  0.5 &  1.4 &  0.8 &  1.0 & 0.332E-09 &  0.3335 $\pm$  0.0065 $\pm$  0.0068 \\
 \hline
  0.1800 &   2.670 &   0.092 &  2.5 &  1.8 &  1.7 &  0.4 &  0.1 & 0.609E-07 &  0.3313 $\pm$  0.0092 $\pm$  0.0117 \\
  0.1800 &   3.480 &   0.114 &  1.9 &  1.2 &  1.1 &  0.3 &  0.1 & 0.341E-07 &  0.3099 $\pm$  0.0061 $\pm$  0.0079 \\
  0.1800 &   4.470 &   0.136 &  1.6 &  0.9 &  0.6 &  0.3 &  0.1 & 0.199E-07 &  0.3143 $\pm$  0.0061 $\pm$  0.0060 \\
  0.1800 &   5.470 &   0.159 &  1.3 &  0.6 &  0.4 &  0.3 &  0.1 & 0.129E-07 &  0.3030 $\pm$  0.0064 $\pm$  0.0047 \\
  0.1800 &   6.860 &   0.178 &  1.1 &  0.4 &  0.5 &  0.3 &  0.2 & 0.792E-08 &  0.3057 $\pm$  0.0056 $\pm$  0.0042 \\
  0.1800 &   8.880 &   0.181 &  1.0 &  0.3 &  0.8 &  0.2 &  0.3 & 0.461E-08 &  0.2991 $\pm$  0.0083 $\pm$  0.0040 \\
  0.1800 &  11.310 &   0.197 &  0.9 &  0.2 &  1.0 &  0.2 &  0.5 & 0.282E-08 &  0.3016 $\pm$  0.0077 $\pm$  0.0045 \\
  0.1800 &  14.970 &   0.184 &  0.2 &  0.3 &  1.3 &  0.2 &  0.6 & 0.163E-08 &  0.3020 $\pm$  0.0060 $\pm$  0.0045 \\
  0.1800 &  19.840 &   0.229 &  0.2 &  0.1 &  1.5 &  0.2 &  0.6 & 0.891E-09 &  0.2933 $\pm$  0.0043 $\pm$  0.0049 \\
  0.1800 &  26.650 &   0.294 &  0.2 &  0.0 &  1.7 &  0.2 &  0.7 & 0.468E-09 &  0.2957 $\pm$  0.0042 $\pm$  0.0056 \\
  0.1800 &  35.130 &   0.378 &  0.2 &  0.6 &  1.9 &  0.4 &  0.8 & 0.252E-09 &  0.2950 $\pm$  0.0052 $\pm$  0.0065 \\
  0.1800 &  45.710 &   0.489 &  0.2 &  0.5 &  2.1 &  0.8 &  1.0 & 0.136E-09 &  0.3133 $\pm$  0.0096 $\pm$  0.0079 \\
 \hline
  0.2250 &   2.900 &   0.079 &  2.5 &  1.8 &  0.9 &  0.3 &  0.1 & 0.397E-07 &  0.2961 $\pm$  0.0263 $\pm$  0.0096 \\
  0.2250 &   3.510 &   0.097 &  2.0 &  1.3 &  0.8 &  0.3 &  0.1 & 0.256E-07 &  0.2950 $\pm$  0.0071 $\pm$  0.0075 \\
  0.2250 &   4.490 &   0.113 &  1.6 &  0.9 &  0.7 &  0.3 &  0.1 & 0.150E-07 &  0.2824 $\pm$  0.0061 $\pm$  0.0057 \\
  0.2250 &   5.460 &   0.129 &  1.4 &  0.6 &  0.6 &  0.3 &  0.1 & 0.974E-08 &  0.2868 $\pm$  0.0066 $\pm$  0.0048 \\
  0.2250 &   6.870 &   0.155 &  1.2 &  0.4 &  0.6 &  0.3 &  0.1 & 0.591E-08 &  0.2807 $\pm$  0.0056 $\pm$  0.0039 \\
  0.2250 &   8.850 &   0.159 &  1.0 &  0.3 &  0.6 &  0.2 &  0.2 & 0.345E-08 &  0.2785 $\pm$  0.0074 $\pm$  0.0035 \\
  0.2250 &  11.350 &   0.162 &  0.9 &  0.2 &  0.7 &  0.2 &  0.5 & 0.206E-08 &  0.2837 $\pm$  0.0076 $\pm$  0.0037 \\
  0.2250 &  15.020 &   0.150 &  0.3 &  0.2 &  1.0 &  0.3 &  0.6 & 0.118E-08 &  0.2564 $\pm$  0.0056 $\pm$  0.0032 \\
  0.2250 &  19.860 &   0.185 &  0.3 &  0.1 &  1.3 &  0.2 &  0.6 & 0.652E-09 &  0.2641 $\pm$  0.0041 $\pm$  0.0039 \\
  0.2250 &  26.710 &   0.238 &  0.3 &  0.1 &  1.7 &  0.2 &  0.6 & 0.342E-09 &  0.2631 $\pm$  0.0040 $\pm$  0.0047 \\
  0.2250 &  35.280 &   0.305 &  0.3 &  0.6 &  2.0 &  0.3 &  0.6 & 0.185E-09 &  0.2547 $\pm$  0.0045 $\pm$  0.0057 \\
  0.2250 &  46.550 &   0.401 &  0.3 &  0.5 &  2.4 &  0.5 &  0.8 & 0.978E-10 &  0.2554 $\pm$  0.0073 $\pm$  0.0069 \\
 \hline
  0.2750 &   3.750 &   0.085 &  1.9 &  1.1 &  1.7 &  0.3 &  0.1 & 0.171E-07 &  0.2679 $\pm$  0.0117 $\pm$  0.0074 \\
  0.2750 &   4.470 &   0.101 &  1.5 &  0.7 &  1.3 &  0.3 &  0.1 & 0.113E-07 &  0.2695 $\pm$  0.0079 $\pm$  0.0058 \\
  0.2750 &   5.470 &   0.109 &  1.3 &  0.5 &  1.0 &  0.3 &  0.1 & 0.727E-08 &  0.2602 $\pm$  0.0074 $\pm$  0.0045 \\
  0.2750 &   6.890 &   0.129 &  1.1 &  0.3 &  0.7 &  0.3 &  0.1 & 0.437E-08 &  0.2489 $\pm$  0.0059 $\pm$  0.0034 \\
  0.2750 &   8.810 &   0.156 &  0.9 &  0.1 &  0.6 &  0.2 &  0.1 & 0.254E-08 &  0.2470 $\pm$  0.0083 $\pm$  0.0028 \\
  0.2750 &  11.420 &   0.138 &  0.8 &  0.0 &  0.7 &  0.2 &  0.4 & 0.148E-08 &  0.2370 $\pm$  0.0078 $\pm$  0.0028 \\
  0.2750 &  14.860 &   0.145 &  0.8 &  0.0 &  0.8 &  0.2 &  0.5 & 0.852E-09 &  0.2400 $\pm$  0.0076 $\pm$  0.0030 \\
  0.2750 &  19.860 &   0.153 &  0.3 &  0.1 &  0.9 &  0.2 &  0.6 & 0.469E-09 &  0.2301 $\pm$  0.0044 $\pm$  0.0027 \\
  0.2750 &  26.720 &   0.195 &  0.3 &  0.3 &  1.2 &  0.2 &  0.6 & 0.247E-09 &  0.2239 $\pm$  0.0040 $\pm$  0.0032 \\
  0.2750 &  35.340 &   0.253 &  0.3 &  0.4 &  1.8 &  0.2 &  0.6 & 0.133E-09 &  0.2268 $\pm$  0.0046 $\pm$  0.0044 \\
  0.2750 &  46.960 &   0.332 &  0.3 &  0.5 &  2.4 &  0.4 &  0.7 & 0.701E-10 &  0.2299 $\pm$  0.0069 $\pm$  0.0059 \\
  0.2750 &  59.800 &   0.417 &  0.4 &  0.5 &  2.9 &  0.6 &  1.3 & 0.401E-10 &  0.1675 $\pm$  0.0179 $\pm$  0.0056 \\
 \hline
 \end{tabular}
 \end{table}
 \newpage
\begin{table}[t]                                                                                                                   
\begin{tabular}{|c|c|c||c|c|c|c|c||c|r|}\hline
$x$ & $Q^2$ & $y$ & $E$ & $E^{\prime}$ & $AC$ & $RC$ & $RE$& $d^2\sigma^{meas}/dxdQ^2$ &
$F^d_2\;\pm\Delta{F^{stat}_2}\;\pm\Delta{F^{syst}_2}$ \\
  & $[{\rm GeV^2}]$ & &  [\%]  &  [\%]  &  [\%]  &  [\%]  &  [\%]  & $[{\rm b\cdot GeV^{-2}}]$ & \\ \hline
 \hline
  0.3500 &   4.610 &   0.082 &  1.0 &  0.1 &  4.4 &  0.3 &  0.1 & 0.728E-08 &  0.2127 $\pm$  0.0081 $\pm$  0.0095 \\
  0.3500 &   5.450 &   0.097 &  0.8 &  0.2 &  3.3 &  0.2 &  0.1 & 0.484E-08 &  0.2217 $\pm$  0.0069 $\pm$  0.0076 \\
  0.3500 &   6.960 &   0.106 &  0.6 &  0.4 &  2.1 &  0.2 &  0.1 & 0.279E-08 &  0.2160 $\pm$  0.0049 $\pm$  0.0047 \\
  0.3500 &   8.810 &   0.125 &  0.5 &  0.5 &  1.2 &  0.2 &  0.1 & 0.164E-08 &  0.2016 $\pm$  0.0057 $\pm$  0.0028 \\
  0.3500 &  11.320 &   0.150 &  0.4 &  0.5 &  0.7 &  0.2 &  0.2 & 0.937E-09 &  0.1831 $\pm$  0.0073 $\pm$  0.0019 \\
  0.3500 &  14.810 &   0.126 &  0.3 &  0.6 &  0.9 &  0.2 &  0.6 & 0.534E-09 &  0.1680 $\pm$  0.0052 $\pm$  0.0021 \\
  0.3500 &  20.200 &   0.129 &  0.3 &  0.6 &  1.2 &  0.2 &  0.6 & 0.278E-09 &  0.1782 $\pm$  0.0038 $\pm$  0.0027 \\
  0.3500 &  26.640 &   0.155 &  0.4 &  0.6 &  1.4 &  0.2 &  0.6 & 0.153E-09 &  0.1677 $\pm$  0.0028 $\pm$  0.0028 \\
  0.3500 &  35.420 &   0.201 &  0.4 &  0.6 &  1.5 &  0.1 &  0.5 & 0.821E-10 &  0.1698 $\pm$  0.0030 $\pm$  0.0030 \\
  0.3500 &  46.630 &   0.259 &  0.4 &  0.5 &  1.6 &  0.2 &  0.5 & 0.445E-10 &  0.1633 $\pm$  0.0038 $\pm$  0.0031 \\
  0.3500 &  61.200 &   0.337 &  0.5 &  0.5 &  2.6 &  0.4 &  1.2 & 0.239E-10 &  0.1556 $\pm$  0.0103 $\pm$  0.0046 \\
 \hline
  0.5000 &   5.720 &   0.071 &  2.2 &  3.3 &  9.4 &  0.2 &  0.1 & 0.191E-08 &  0.1148 $\pm$  0.0100 $\pm$  0.0117 \\
  0.5000 &   6.920 &   0.086 &  1.9 &  3.1 &  7.0 &  0.2 &  0.1 & 0.118E-08 &  0.1263 $\pm$  0.0047 $\pm$  0.0100 \\
  0.5000 &   8.880 &   0.093 &  1.8 &  3.0 &  4.4 &  0.2 &  0.1 & 0.641E-09 &  0.1053 $\pm$  0.0039 $\pm$  0.0060 \\
  0.5000 &  11.270 &   0.105 &  1.9 &  3.0 &  2.5 &  0.2 &  0.2 & 0.362E-09 &  0.1027 $\pm$  0.0044 $\pm$  0.0045 \\
  0.5000 &  14.370 &   0.133 &  1.4 &  2.6 &  1.5 &  0.1 &  0.2 & 0.206E-09 &  0.1085 $\pm$  0.0061 $\pm$  0.0036 \\
  0.5000 &  20.030 &   0.110 &  2.1 &  3.3 &  2.0 &  0.2 &  0.7 & 0.999E-10 &  0.0876 $\pm$  0.0035 $\pm$  0.0039 \\
  0.5000 &  27.190 &   0.116 &  0.9 &  0.7 &  2.8 &  0.1 &  0.6 & 0.509E-10 &  0.0848 $\pm$  0.0020 $\pm$  0.0026 \\
  0.5000 &  35.480 &   0.143 &  0.8 &  0.6 &  3.3 &  0.1 &  0.5 & 0.282E-10 &  0.0823 $\pm$  0.0016 $\pm$  0.0029 \\
  0.5000 &  46.620 &   0.183 &  0.8 &  0.6 &  3.6 &  0.1 &  0.5 & 0.153E-10 &  0.0805 $\pm$  0.0020 $\pm$  0.0030 \\
  0.5000 &  62.310 &   0.243 &  0.8 &  0.5 &  3.7 &  0.2 &  1.1 & 0.790E-11 &  0.0723 $\pm$  0.0047 $\pm$  0.0029 \\
 \hline
 \end{tabular}
 \end{table}

\clearpage


\begin{thebibliography}{99}
\bibitem{f2nmc} NMC, P.~Amaudruz et al., Phys. Lett.{\bf B 295} (1992) 159.
\bibitem{longratio} NMC, P.~Amaudruz et al., Nucl. Phys. {\bf B 371} (1992) 3.
\bibitem{apparatus} M. van der Heijden, Ph.D. Thesis, University of Amsterdam
(1991);\\ I.G. Bird, Ph.D. Thesis, Free University, Amsterdam (1992);\\
A. Br\"{u}ll, Ph.D. Thesis, Freiburg University (1992);\\
T. Granier, Ph.D. Thesis, Universit\'{e} Pierre et Marie Curie, Paris (1994);\\
P. Bj\"{o}rkholm, Ph.D. Thesis, Uppsala University (1995).
\bibitem{t10} R.P.~Mount, Nucl. Instrum. Methods {\bf 187} (1981) 401.
\bibitem{bcs} M.~Arneodo, Ph.D. Thesis, Princeton University (1992).
\bibitem{15param} A. Milsztajn et al., Z. Phys. {\bf C 49} (1991) 527.
\bibitem{rslac} L.W.~Whitlow et al., Phys. Lett. {\bf B 250} (1990) 193.
\bibitem{radcor} A.A.~Akhundov et al., Sov. J. Nucl. Phys. {\bf 26} (1977) 660;
44 (1986) 988;\\
JINR-Dubna preprints E2-10147 (1976), E2-10205 (1976),
E2-86-104 (1986);\\
D.~Bardin and N.~Shumeiko, Sov. J .Nucl. Phys. {\bf 29} (1979) 499.
\bibitem{slac92} L.W. Whitlow et al., SLAC, Phys.Lett {\bf B 282} (1992) 475.
\bibitem{BCDMS} BCDMS Collab., A.C. Benvenuti et al., Phys. Lett. {\bf B
233} (1989) 485;\\
BCDMS Collab., A.C. Benvenuti et al., Phys. Lett. {\bf B
237} (1990) 592.
\bibitem{ref:E665} E665 Collab., presented at the 30th
Rencontres de Moriond, March 1995;
A.V. Kotwal, Ph.D. Thesis, Harvard University (1995).
\bibitem{ref:h1} H1 Collab., T. Ahmed et al., Nucl. Phys.
{\bf B 439} (1995) 471.
\bibitem{ref:zeus} ZEUS Collab., M. Derrick et al., Z. Phys.
{\bf C 65} (1995) 627.

\end{thebibliography}
\end{document}